\RequirePackage{lineno}
\documentclass[aps,prd,showpacs,superscriptaddress,twocolumn]{revtex4}  
\usepackage{graphicx}  
\usepackage{epsf}  
\usepackage{dcolumn}   

\usepackage{setspace}  

\usepackage{bm}        
\usepackage{ulem}      
\usepackage{color}     %
\usepackage{amssymb}   
\usepackage{multirow}
\usepackage{epstopdf}  

\newcommand {\DeltaR}        {\mbox{$\Delta R$}}

\newcommand {\eeee}          {\mbox{$eeee$}}
\newcommand {\eejj}          {\mbox{$eejj$}}
\newcommand {\eemm}          {\mbox{$ee\mu\mu$}}
\newcommand {\et}            {\mbox{${E_T}$}}
\newcommand {\Et}            {\ensuremath{E_{T}}}

\newcommand {\gev}           {\mbox{$\,\rm GeV$}}
\newcommand {\gevc}          {\mbox{$\,\rm GeV\!/{\it c}$}}
\newcommand {\gevcsq}        {\mbox{$\,\rm GeV\!/{\it c}^2$}}
\newcommand {\gt}            {\mbox{$>$}}

\newcommand {\invfb}         {\mbox{\,fb$^{-1}$}}

\newcommand {\lljj}          {\mbox{$\ell\ell jj$}}
\newcommand {\llmet}         {\mbox{$\ell\ell + {\not \! E}_{T}$}}
\newcommand {\lt}            {\mbox{$<$}}
\newcommand {\lumi}          {\mbox{\rm (lumi.)}}
\newcommand {\luminosity}    {$6$\,fb$^{-1}$}
\newcommand {\mgstar}        {\mbox{$M_{G^*}=325\,{\rm GeV}\!/{\it c}^2$}}
\newcommand {\mjj}           {\mbox{$M_{jj}$}}
\newcommand {\mll}           {\mbox{$M_{\ell\ell}$}}
\newcommand {\mllll}         {\mbox{$M_{\ell\ell\ell\ell}$}}
\newcommand {\mlljj}         {\mbox{$M_{\ell\ell jj}$}}
\newcommand {\met}           {\mbox{${\not \! E}_{T}$}}
\newcommand {\metx}          {\mbox{${\not \! E}_{x}$}}
\newcommand {\mety}          {\mbox{${\not \! E}_{y}$}}
\newcommand {\mmjj}          {\mbox{$\mu\mu jj$}}
\newcommand {\mmmm}          {\mbox{$\mu\mu\mu\mu$}}
\newcommand {\mvis}          {\mbox{$M_{ZZ}^{\rm vis}$}}
\newcommand {\mzz}           {\mbox{$M_{ZZ}$}}

\newcommand {\ppbar}         {\mbox{$p\bar{p}$}}
\newcommand {\pb}            {\mbox{\rm\,pb}}
\newcommand {\pt}            {\mbox{$p_{T}$}}

\newcommand {\stat}          {\mbox{\rm (stat.)}}
\newcommand {\statsys}       {\mbox{\rm (stat.+syst.)}}
\newcommand {\syst}          {\mbox{\rm (syst.)}}
\newcommand {\tevcsq}        {\mbox{$\,\rm TeV\!/{\it c}^2$}}
\newcommand {\ttbar}         {\mbox{$t\bar{t}$}}

\newcommand {\wenu}          {\mbox{$W^{\pm} \rightarrow e^{\pm} \nu$}}

\newcommand {\ww}            {\mbox{$WW$}}
\newcommand {\wz}            {\mbox{$WZ$}}
\newcommand {\zgamma}        {\mbox{$Z / \gamma^*$}}
\newcommand {\z}             {\mbox{$Z$}}
\newcommand {\zee}           {\mbox{$Z \rightarrow e^+ e^-$}}
\newcommand {\zll}           {\mbox{$Z \rightarrow \ell^+ \ell^-$}}
\newcommand {\zjj}           {\mbox{$Z \to j j$}}

\newcommand {\zmumu}         {\mbox{$Z \rightarrow \mu^+ \mu^-$}}

\newcommand {\zz}            {\mbox{$ZZ$}}
\newcommand {\zzllll}        {\mbox{$ZZ \to \ell^+ \ell^- \ell^+ \ell^-$}}
\newcommand {\zzllnn}        {\mbox{$ZZ \to \ell^+ \ell^- \nu \nu$}}
\newcommand {\zzlljj}        {\mbox{$ZZ \to \ell^+ \ell^- j j$}}

\input{latex_input.tmp}

\newcommand {\plots} {.}   

\begin{document}

\topmargin 0.1in

\newif \ifpublic
\publictrue



\hspace{5.2in}  
\mbox{\hfill Fermilab-PUB-11-613-E}

\title{Search for high-mass resonances decaying into \zz\ in \ppbar\ collisions at $\sqrt{s}=1.96$\,TeV}

\affiliation{Institute of Physics, Academia Sinica, Taipei, Taiwan 11529, Republic of China}
\affiliation{Argonne National Laboratory, Argonne, Illinois 60439, USA}
\affiliation{University of Athens, 157 71 Athens, Greece}
\affiliation{Institut de Fisica d'Altes Energies, ICREA, Universitat Autonoma de Barcelona, E-08193, Bellaterra (Barcelona), Spain}
\affiliation{Baylor University, Waco, Texas 76798, USA}
\affiliation{Istituto Nazionale di Fisica Nucleare Bologna, $^{ee}$University of Bologna, I-40127 Bologna, Italy}
\affiliation{University of California, Davis, Davis, California 95616, USA}
\affiliation{University of California, Los Angeles, Los Angeles, California 90024, USA}
\affiliation{Instituto de Fisica de Cantabria, CSIC-University of Cantabria, 39005 Santander, Spain}
\affiliation{Carnegie Mellon University, Pittsburgh, Pennsylvania 15213, USA}
\affiliation{Enrico Fermi Institute, University of Chicago, Chicago, Illinois 60637, USA}
\affiliation{Comenius University, 842 48 Bratislava, Slovakia; Institute of Experimental Physics, 040 01 Kosice, Slovakia}
\affiliation{Joint Institute for Nuclear Research, RU-141980 Dubna, Russia}
\affiliation{Duke University, Durham, North Carolina 27708, USA}
\affiliation{Fermi National Accelerator Laboratory, Batavia, Illinois 60510, USA}
\affiliation{University of Florida, Gainesville, Florida 32611, USA}
\affiliation{Laboratori Nazionali di Frascati, Istituto Nazionale di Fisica Nucleare, I-00044 Frascati, Italy}
\affiliation{University of Geneva, CH-1211 Geneva 4, Switzerland}
\affiliation{Glasgow University, Glasgow G12 8QQ, United Kingdom}
\affiliation{Harvard University, Cambridge, Massachusetts 02138, USA}
\affiliation{Division of High Energy Physics, Department of Physics, University of Helsinki and Helsinki Institute of Physics, FIN-00014, Helsinki, Finland}
\affiliation{University of Illinois, Urbana, Illinois 61801, USA}
\affiliation{The Johns Hopkins University, Baltimore, Maryland 21218, USA}
\affiliation{Institut f\"{u}r Experimentelle Kernphysik, Karlsruhe Institute of Technology, D-76131 Karlsruhe, Germany}
\affiliation{Center for High Energy Physics: Kyungpook National University, Daegu 702-701, Korea; Seoul National University, Seoul 151-742, Korea; Sungkyunkwan University, Suwon 440-746, Korea; Korea Institute of Science and Technology Information, Daejeon 305-806, Korea; Chonnam National University, Gwangju 500-757, Korea; Chonbuk National University, Jeonju 561-756, Korea}
\affiliation{Ernest Orlando Lawrence Berkeley National Laboratory, Berkeley, California 94720, USA}
\affiliation{University of Liverpool, Liverpool L69 7ZE, United Kingdom}
\affiliation{University College London, London WC1E 6BT, United Kingdom}
\affiliation{Centro de Investigaciones Energeticas Medioambientales y Tecnologicas, E-28040 Madrid, Spain}
\affiliation{Massachusetts Institute of Technology, Cambridge, Massachusetts 02139, USA}
\affiliation{Institute of Particle Physics: McGill University, Montr\'{e}al, Qu\'{e}bec, Canada H3A~2T8; Simon Fraser University, Burnaby, British Columbia, Canada V5A~1S6; University of Toronto, Toronto, Ontario, Canada M5S~1A7; and TRIUMF, Vancouver, British Columbia, Canada V6T~2A3}
\affiliation{University of Michigan, Ann Arbor, Michigan 48109, USA}
\affiliation{Michigan State University, East Lansing, Michigan 48824, USA}
\affiliation{Institution for Theoretical and Experimental Physics, ITEP, Moscow 117259, Russia}
\affiliation{University of New Mexico, Albuquerque, New Mexico 87131, USA}
\affiliation{The Ohio State University, Columbus, Ohio 43210, USA}
\affiliation{Okayama University, Okayama 700-8530, Japan}
\affiliation{Osaka City University, Osaka 588, Japan}
\affiliation{University of Oxford, Oxford OX1 3RH, United Kingdom}
\affiliation{Istituto Nazionale di Fisica Nucleare, Sezione di Padova-Trento, $^{ff}$University of Padova, I-35131 Padova, Italy}
\affiliation{University of Pennsylvania, Philadelphia, Pennsylvania 19104, USA}
\affiliation{Istituto Nazionale di Fisica Nucleare Pisa, $^{gg}$University of Pisa, $^{hh}$University of Siena and $^{ii}$Scuola Normale Superiore, I-56127 Pisa, Italy}
\affiliation{University of Pittsburgh, Pittsburgh, Pennsylvania 15260, USA}
\affiliation{Purdue University, West Lafayette, Indiana 47907, USA}
\affiliation{University of Rochester, Rochester, New York 14627, USA}
\affiliation{The Rockefeller University, New York, New York 10065, USA}
\affiliation{Istituto Nazionale di Fisica Nucleare, Sezione di Roma 1, $^{jj}$Sapienza Universit\`{a} di Roma, I-00185 Roma, Italy}
\affiliation{Rutgers University, Piscataway, New Jersey 08855, USA}
\affiliation{Texas A\&M University, College Station, Texas 77843, USA}
\affiliation{Istituto Nazionale di Fisica Nucleare Trieste/Udine, I-34100 Trieste, $^{kk}$University of Udine, I-33100 Udine, Italy}
\affiliation{University of Tsukuba, Tsukuba, Ibaraki 305, Japan}
\affiliation{Tufts University, Medford, Massachusetts 02155, USA}
\affiliation{University of Virginia, Charlottesville, Virginia 22906, USA}
\affiliation{Waseda University, Tokyo 169, Japan}
\affiliation{Wayne State University, Detroit, Michigan 48201, USA}
\affiliation{University of Wisconsin, Madison, Wisconsin 53706, USA}
\affiliation{Yale University, New Haven, Connecticut 06520, USA}

\author{T.~Aaltonen}
\affiliation{Division of High Energy Physics, Department of Physics, University of Helsinki and Helsinki Institute of Physics, FIN-00014, Helsinki, Finland}
\author{B.~\'{A}lvarez~Gonz\'{a}lez$^z$}
\affiliation{Instituto de Fisica de Cantabria, CSIC-University of Cantabria, 39005 Santander, Spain}
\author{S.~Amerio}
\affiliation{Istituto Nazionale di Fisica Nucleare, Sezione di Padova-Trento, $^{ff}$University of Padova, I-35131 Padova, Italy}
\author{D.~Amidei}
\affiliation{University of Michigan, Ann Arbor, Michigan 48109, USA}
\author{A.~Anastassov$^x$}
\affiliation{Fermi National Accelerator Laboratory, Batavia, Illinois 60510, USA}
\author{A.~Annovi}
\affiliation{Laboratori Nazionali di Frascati, Istituto Nazionale di Fisica Nucleare, I-00044 Frascati, Italy}
\author{J.~Antos}
\affiliation{Comenius University, 842 48 Bratislava, Slovakia; Institute of Experimental Physics, 040 01 Kosice, Slovakia}
\author{G.~Apollinari}
\affiliation{Fermi National Accelerator Laboratory, Batavia, Illinois 60510, USA}
\author{J.A.~Appel}
\affiliation{Fermi National Accelerator Laboratory, Batavia, Illinois 60510, USA}
\author{T.~Arisawa}
\affiliation{Waseda University, Tokyo 169, Japan}
% ...

\collaboration{CDF Collaboration\footnote{With visitors from
$^a$Istituto Nazionale di Fisica Nucleare, Sezione di Cagliari, 09042 Monserrato (Cagliari), Italy,
$^b$University of CA Irvine, Irvine, CA 92697, USA,
$^c$University of CA Santa Barbara, Santa Barbara, CA 93106, USA,
$^d$University of CA Santa Cruz, Santa Cruz, CA 95064, USA,
$^e$Institute of Physics, Academy of Sciences of the Czech Republic, Czech Republic,
$^f$CERN, CH-1211 Geneva, Switzerland,
$^g$Cornell University, Ithaca, NY 14853, USA,
$^h$University of Cyprus, Nicosia CY-1678, Cyprus,
$^i$Office of Science, U.S. Department of Energy, Washington, DC 20585, USA,
$^j$University College Dublin, Dublin 4, Ireland,
$^k$ETH, 8092 Zurich, Switzerland,
$^l$University of Fukui, Fukui City, Fukui Prefecture, Japan 910-0017,
$^m$Universidad Iberoamericana, Mexico D.F., Mexico,
$^n$University of Iowa, Iowa City, IA 52242, USA,
$^o$Kinki University, Higashi-Osaka City, Japan 577-8502,
$^p$Kansas State University, Manhattan, KS 66506, USA,
$^q$Korea University, Seoul, 136-713, Korea,
$^r$University of Manchester, Manchester M13 9PL, United Kingdom,
$^s$Queen Mary, University of London, London, E1 4NS, United Kingdom,
$^t$University of Melbourne, Victoria 3010, Australia,
$^u$Muons, Inc., Batavia, IL 60510, USA,
$^v$Nagasaki Institute of Applied Science, Nagasaki, Japan,
$^w$National Research Nuclear University, Moscow, Russia,
$^x$Northwestern University, Evanston, IL 60208, USA,
$^y$University of Notre Dame, Notre Dame, IN 46556, USA,
$^z$Universidad de Oviedo, E-33007 Oviedo, Spain,
$^{aa}$CNRS-IN2P3, Paris, F-75205 France,
$^{bb}$Texas Tech University, Lubbock, TX 79609, USA,
$^{cc}$Universidad Tecnica Federico Santa Maria, 110v Valparaiso, Chile,
$^{dd}$Yarmouk University, Irbid 211-63, Jordan,
}}
\noaffiliation
\date{\today}
 
\begin{abstract}

  We search for high-mass resonances decaying into \z\ boson pairs
  using data corresponding to 6\invfb\ collected by the CDF experiment in \ppbar\ 
  collisions at $\sqrt{s}=1.96$\,TeV.    
  The search is performed in three distinct final states:
  \zzllll, \zzllnn, and \zzlljj. 
  For a Randall-Sundrum graviton $G^*$, 
  the 95\% CL upper limits on the production cross section 
  times branching ratio to \zz, 
  $\sigma(\ppbar\to G^* \to\zz)$, vary between 0.26\,pb and 0.045\,pb
  in the mass range $300<M_{G^*}<1000\gevcsq$.

\end{abstract}

\pacs{13.85.Rm, 14.70.Hp, 14.70.Kv}

\maketitle

\section{Introduction}

We report the results of a search for high-mass resonances 
decaying to \zz\ in \ppbar\ collisions 
at $\sqrt{s} = 1.96$\,TeV at the Tevatron.
Although 
the decay of the standard model Higgs boson to \zz\ is expected to be beyond 
the sensitivity of the Tevatron experiments \cite{higgsxs},
new physics could affect \zz\ production in different ways.
In models containing large extra dimensions
the \zz\ production cross section is increased through loop corrections 
\cite{ZZ_EXTRA_DIMENSIONS}.
Resonances appearing at high mass such as a 
Randall-Sundrum (RS) graviton \cite{rsgraviton} 
could decay manifestly to two \z\ bosons.
The original RS model predicts Kaluza-Klein excitations 
of the graviton ($G^*$) that decay predominantly 
to a pair of charged leptons or a pair of photons.  Experimental 
searches for such high-mass resonance decays have excluded
RS graviton states up to a mass of around 1\tevcsq\ at 95\% confidence level
for a natural choice of coupling parameter \cite{k_over_mPl},
both at the Tevatron 
and at the LHC \cite{previousresults}.
However, in RS models that have standard model fields propagating in the 
bulk, the $G^*$ couplings to light fermions and photons 
may be heavily suppressed 
so that the dominant decay modes are to \ttbar, Higgs pairs, 
or pairs of heavy bosons \cite{bulk}.
Furthermore, in some models  
the decay to heavy bosons is dominant \cite{lisafitzpatrick}.
Suppression of the couplings to light fermions also results in 
gluon fusion becoming the primary production process. 

The CDF experiment has previously searched for resonances decaying 
to \z\ pairs and excluded RS gravitons with mass up to around 0.5\tevcsq\
at 95\% confidence level \cite{antonio}.
The search described in this paper 
gives improved sensitivity over the previous analysis
through modified event selection, 
the inclusion of extra final states, and the 
addition of more data.
Three final states are examined, corresponding to the different 
\z\ boson decay modes \zzllll, \zzllnn, and \zzlljj, where $\ell$ 
is an electron or muon and $j$ is a hadronic jet.
These three channels have different signal-to-background
ratios and allow an overconstrained search.
The four-lepton final state has 
the smallest background; however, depending on the resonance mass, 
the best single-channel sensitivity is provided by either the 
\zzlljj\ or \zzllnn\ channels. 

The paper is organized as follows: in Section~\ref{det} we introduce 
the CDF detector and trigger system; in Section~\ref{leptonid} we describe 
the reconstruction and identification procedures; then in
Sections~\ref{zz4l}--\ref{zzlljj} we report the search results from 
each of the channels \zzllll, \zzllnn\ and \zzlljj.
Section~\ref{section:limits} gives limits resulting from all three channels
and their combination.


\section{Detector}
\label{det}

The CDF II detector is a general-purpose particle detector, described
in detail elsewhere~\cite{cdftdr}. 
The results reported in this paper use 
information from several detector subsystems for charged lepton and jet 
reconstruction and identification. 

Tracks of charged particles are reconstructed in the silicon system~\cite{silicon} 
and in the 
central tracker~\cite{cot}, which is a drift chamber 
that consists of 96 layers of sense wires grouped into eight `superlayers'.
Superlayers alternate between an axial configuration, with sense wires 
parallel to the colliding beams, and a small-angle stereo configuration.
For high momentum tracks the resolution is
$\sigma_{p_T}/p_T^2\simeq 1.7\times 10^{-3}(\gevc)^{-1}$, 
where $\pt = p \sin\theta$, $p$ being the track momentum and 
$\theta$ the polar angle with respect to the proton beam direction.

The calorimeter is segmented radially into electromagnetic and 
hadronic compartments \cite{cdf_cem_ces,cdf_cha}.
The central calorimeter is split at the center
into two separate barrels and covers the pseudorapidity range $|\eta|<1.1$ 
(where $\eta = -\ln \tan \frac{\theta}{2}$).
Each barrel consists of 24 azimuthal wedges segmented in 
projective towers of 0.1 in $\eta$.
The forward calorimeter segmentation increases from 
0.1 in $\eta$ and $7.5^\circ$ in the azimuthal angle $\phi$ at $\eta=1.1$,
to 
0.5 in $\eta$ and $15^\circ$ in $\phi$ at $\eta=3.6$.
Electron energy resolutions are $13.5\%/\sqrt{E_T}\oplus 2\%$ in the 
central calorimeter and $16\%/\sqrt{E}\oplus 1\%$ in the 
forward calorimeters, where $\et = E \sin\theta$.
The electromagnetic calorimeters incorporate shower maximum 
detectors that are used to measure shower profiles 
with spatial resolution of around 2\,mm.

Dedicated muon detectors~\cite{cdf_muon_system} are mounted around 
the calorimeters, providing coverage for $|\eta|\lesssim 1.5$.
Luminosity is measured by a hodoscopic system of 
Cherenkov counters~\cite{clc}.

CDF has a three-level online trigger system.
The data used in this measurement were collected using 
inclusive high-\pt\ electron and muon triggers, 
and a two-electron trigger.
The single-lepton triggers select events that have
electron or muon candidates with $\pt\geq 18$\gevc\
and ${\rm |\eta|} \lesssim 1.0$ \cite{drell-yan}, and the two-electron trigger
uses only calorimeter information and 
allows electron candidates above the same \pt\ threshold 
anywhere in the detector.
The data correspond
to an integrated luminosity of \luminosity\ collected between 
February 2002 and February 2010.


\section{Reconstruction and Identification}
\label{leptonid}

In this section we discuss lepton reconstruction and identification, 
and reconstruction of jets and missing transverse energy.

\subsection{Leptons}

Decays of a heavy resonance to \zz, where at least one of the \z\ 
bosons decays leptonically, result in a wide lepton energy spectrum.
Any inefficiency in lepton reconstruction and identification is raised 
to the fourth power in the \zzllll\ channel.  Thus, keeping 
efficiency high while maintaining stringent background rejection is equally important 
for $\pt\sim 20$\gevc\ and for $\pt>100$\gevc.
To this end, this analysis incorporates several refinements in the 
offline reconstruction and identification of electron and muon candidates.
Studies were performed on inclusive \zll\ candidates and on events 
containing one lepton plus two additional tracks having $\pt > 10\gevc$, 
and this latter data set 
was fully reprocessed for the \zzllll\ analysis.

First we describe the elements of the lepton selection that are standard to CDF.
Electron candidates consist of a calorimeter cluster matched to a well-reconstructed
track.  Candidates are required to be within the fiducial region of 
 the shower maximum 
detectors and have a shower that is mostly contained in the electromagnetic 
compartment of the calorimeter, with a shower shape that is consistent with 
test beam expectation \cite{run1wz}.
For candidates reconstructed in the central part of the detector 
($|\eta|<1.1$), the matched track must have 
\pt$>$10\gevc, pass through all layers of the central tracker, 
and have a fit $\chi^2/{\rm d.o.f.}<3$.
Candidates reconstructed in the forward part of the detector,
$1.13<|\eta|<2.8$, must either have a track in the central tracker, 
or a track in the silicon system with $\geq5$ hits.

A muon candidate is reconstructed from a track in the central tracker
pointing to track segments in the muon chambers.
Muon track trajectories must be such that at least 30 central tracker 
hits would be expected geometrically, and at least 60\% of those must be found. 
Tracks pointing forward that have fewer 
than three central tracker segments must also have at least five 
$r-\phi$ hits in the silicon tracking system.
Muon energy deposition must be consistent with that of a 
minimally-ionizing particle.
We also consider minimally-ionizing tracks 
that have no track segments in the muon systems 
as muon candidates.

Electron and muon candidates are required to have 
$\et>15$\gev\ and $\pt>15$\gevc\ respectively.
In addition, one of the lepton candidates in each event is required to 
have $\et>20$\gev\ (electrons) or $\pt>20$\gevc\ (muons), and to 
pass more restrictive quality requirements.
These extra requirements are that the lepton track must have 
at least three segments reconstructed in the axial 
superlayers and three in the stereo superlayers; and the track 
of a muon candidate must also be well-matched to a track segment
reconstructed in the muon system.

\begin{figure}[hbt]
    \vspace{0.0in}
    \includegraphics[width=0.5\textwidth, clip=true, viewport=.1in .0in 7.7in 3.3in]
    {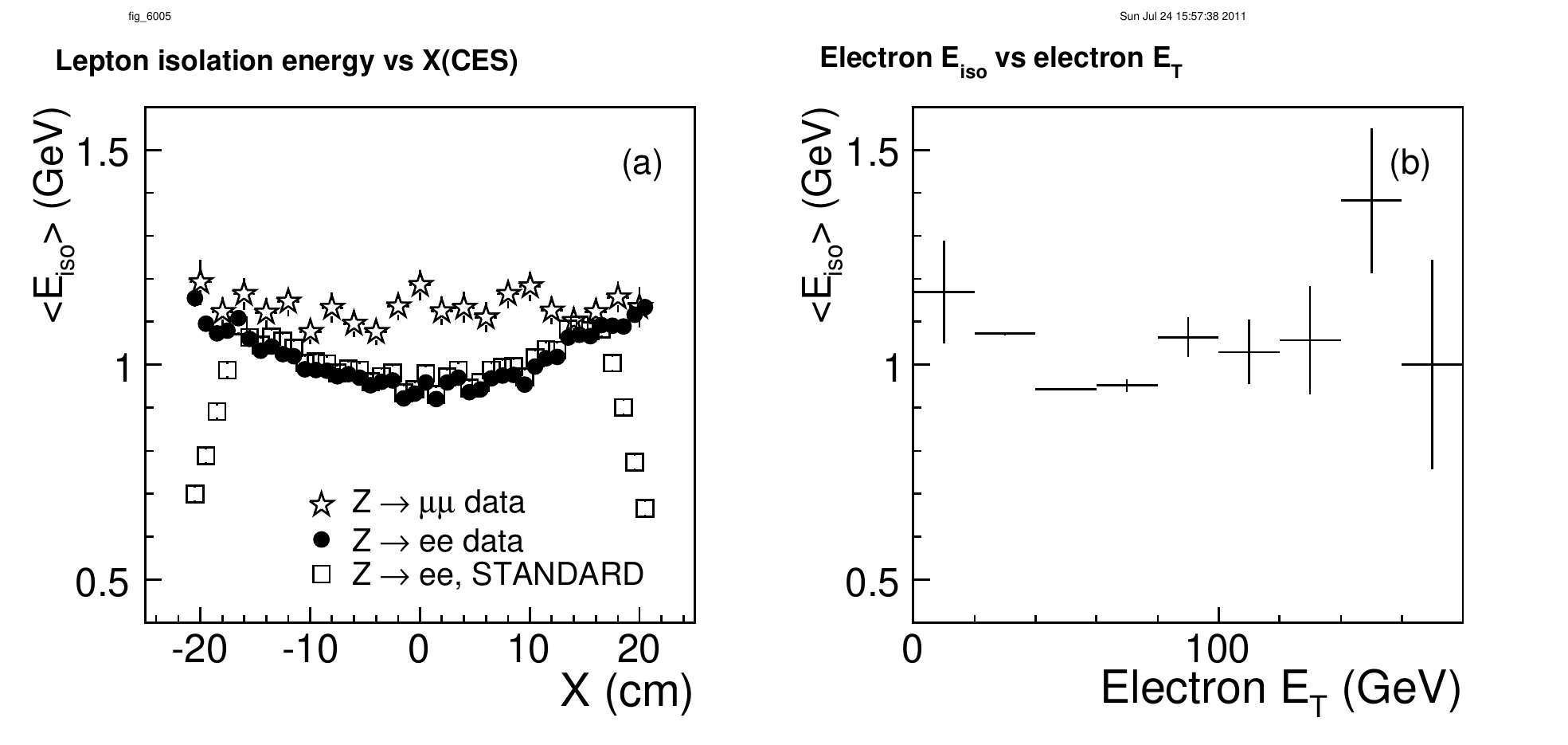}
    \caption[]{
      (a) Corrected isolation energy across the calorimeter
      wedge coordinate $X$ in \zee\ (new correction: solid circles; standard
      correction: open squares) and \zmumu\ (new correction: open
      stars) events; 
      (b) average calorimeter isolation energy as a function of 
      electron \et\ in \zee\ events.}
    \label{fig:ces_dist}
  \end{figure}

The first refinement in lepton selection is in the isolation 
requirement made on all lepton candidates. 
The `isolation energy' is the amount of energy reconstructed in a cone
of $\DeltaR < 0.4$ around a lepton candidate, where
$\DeltaR = \sqrt{(\Delta\eta)^2+(\Delta\phi)^2}$. 
In computing the isolation energy, we refine the treatment of 
energy leakage across calorimeter cell boundaries.
In the central calorimeter, electron clusters include energy 
depositions from only a single wedge in $\phi$.
As each calorimeter tower is read out from different $\phi$ sides
by two photomultiplier tubes, the relative
heights of the pulses locate the energy deposition in $\phi$. 
Locating the 
center of the energy depositions
in towers neighboring the electron cluster allows us 
to estimate the leakage, and correct the isolation 
energy variable 
event-by-event, rather than by applying an average correction.
The correction method is validated by examining the isolation 
energy as a function of shower position in the calorimeter cell, 
which is found to be more uniform than under application of 
the standard average correction,
as shown in Fig.~\ref{fig:ces_dist}(a).  Muons are not expected 
to result in energy leakage; their isolation energy is also shown in 
Fig.~\ref{fig:ces_dist}(a) as validation of the method.
The average isolation energy should depend on the 
instantaneous luminosity but not on the lepton \et, 
and its uniformity in lepton \et\ is confirmed by Fig.~\ref{fig:ces_dist}(b).
All electron and muon candidates are therefore required to be isolated 
in the calorimeter by limiting the isolation energy to be below 4\gev.
Cutting on isolation energy, rather than requiring the standard ratio of 
isolation energy to lepton momentum to be $<0.1$ \cite{run1wz}, increases 
the acceptance for \zzllll\ events by 4\%.

For the \zzllll\ analysis, 
events have been reconstructed with an updated version of the 
CDF tracking code
that gives improved pattern recognition at high luminosities.
The updated version includes an extra algorithm to associate 
hits in the central tracker with silicon-only 
tracks from electron candidates in the forward
region of the detector.  Adding extra hits on to these tracks 
improves the robustness of forward electron charge identification.

Use of an improved reconstruction algorithm in the central shower maximum 
detector gives better separation between showers generated by electron tracks 
and showers produced by bremsstrahlung photons.
Matching tracks to the showers they initiate in both  
coordinate and energy improves hadron rejection and allows the inclusion of 
electron candidates that lose a significant amount of energy through 
bremsstrahlung.
The improved background rejection allows the relaxation of other 
standard electron identification requirements and, overall, 
the selection efficiency is increased by around 9\% per electron.

Electrons reconstructed in the edge $\phi$-rings of the calorimeter on either 
side of the gap between the central and forward detectors
are generally excluded from analysis.
They are included here, after 
verification that they have energy resolution 
comparable with electrons reconstructed in the bulk of the detectors, 
and are well-modeled in the simulation.  
This increases electron acceptance by around 10\% per electron.

The combined effect of the refinements described above is to increase 
lepton acceptance without increasing fake lepton backgrounds, 
as measured by jet-to-lepton fake rates in inclusive jet datasets.
The lepton selection used for this analysis 
is validated by measuring inclusive \zll\ 
cross-sections and separating events by calorimeter region and 
muon system.  
We verify that for each subset of events the measurement 
is stable in time, 
and combining all channels we measure 
$\sigma(\ppbar \to Z)\times Br(\zll)= 247\pm 6\,\statsys \pm 15\,\lumi$\pb, 
consistent with CDF's measurement \cite{drell-yan}.

\subsection{Jets and \met}

Jets are reconstructed as clustered energy depositions in the calorimeter
using a fixed cone clustering algorithm with cone size $\Delta R=0.4$ \cite{jetclustering}.
Jet energies are corrected for $\eta$-dependent calorimeter 
response and for multiple
interactions \cite{jetcorr}. We consider jets having $\et>20$\gev.

The missing transverse energy (\met) is defined as the sum over 
calorimeter tower energies $\vec{\met} = -\sum_i E_T^i{\bf n_i}$, 
where ${\bf n_i}$ is the unit vector in the transverse plane 
that points to calorimeter tower $i$.  The \met\ is 
adjusted to account for the energy corrections made to 
reconstructed jets, and for muons identified in the event.
As neutrinos pass through the detector without depositing energy,
large \met\ in an event can imply the presence of high-energy neutrinos.


\section{\zzllll\ channel}
\label{zz4l}

The first search channel is \zzllll.
We select events with four candidate charged leptons, which may be 
electrons or muons. 
At least two of the four must have $\et>20\gev$ for electron candidates 
($\pt>20\gevc$ for muon candidates) and 
pass the more restrictive lepton selection; and in order 
to have the trigger efficiency well-defined, 
at least one must satisfy the trigger requirements. 

Leptons of the same flavor are paired to form \z\ candidates, 
seeded by a lepton that passes the tighter selection.
In the case of four-electron or four-muon candidates, the pairings 
that minimize the $\chi^2$ of the \zz\ hypothesis are chosen:
$$
\chi^2 ~=~ (M_{12} - M_Z)^2/\sigma_M^2 + (M_{34} - M_Z)^2/\sigma_M^2,
$$
where $M_{12}$ and $M_{34}$ are the masses of the lepton pairs, 
$\sigma_M = 3$ \gevcsq\ approximates experimental resolution in \mll\  
for both electron and muon decays, 
and $M_Z$ is the mass of the \z\ boson.

We find ten events that pass the four-lepton selection.  
In all of these events the number of leptons of the same flavor 
is even.
The best pairings of the ten candidate events are all oppositely-charged.
To minimize the effect of \zgamma\ interference, both \z\ boson candidates 
are required to be within 15\,\gevcsq\ of the \z\ pole, 
$76<M_{\ell\ell}<106$\,\gevcsq.  Following this requirement, eight event candidates remain:
two events have four reconstructed electrons (\eeee), three have two electrons 
and two muons (\eemm), and the remaining three have four reconstructed muons (\mmmm).
The two events that fail the \z\ mass requirement both have one 
\z\ candidate with invariant mass below 60\gevcsq.

\begin{figure}[h!]
  \begin{center}
    \includegraphics[width=0.45\textwidth, clip=true, viewport=1.2in 9.5in 8.in 10.9in]
    {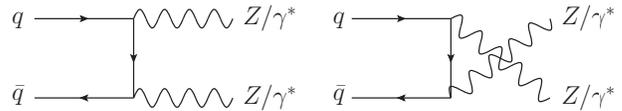}
    \caption[]{ 
	Lowest-order standard model \zz\ production.
    }
    \label{fig:feyn}
  \end{center}
\end{figure}

We use the selected events to measure the 
$\ppbar\to\zz$ production cross section. 
On- and off-shell \zz\ production, as shown in Fig.\,\ref{fig:feyn}, 
followed by \z\ boson decays to charged 
leptons, is the only lowest-order standard model process that results 
in a final 
state with four high-$p_T$ leptons produced in the primary interaction. 
The background in this channel thus comes only from misidentification. 
The main contributions are:
$\ppbar\to\wz\ +$~jet with a jet misidentified as a lepton;
$\ppbar\to\z + 2$~jets with both jets misidentified as leptons;
and $\ppbar\to\z + \gamma\ +$~jet with both the photon and the jet 
 misidentified as electrons.
The contribution from $\ttbar$ production is an order of magnitude 
smaller than that of \wz\ production.
As a result of the \mll \gt 76 \gevcsq\ requirement, 
the contribution of $Z \to \tau\tau$ decays is negligible.

\label{section:fakerates}
The {\sc pythia} event generator \cite{pythia} and the full CDF 
detector simulation \cite{cdfsim}
are used to simulate kinematics of these processes
and photon-to-lepton misidentification. 
Jet-to-lepton misidentification rates are 
measured in inclusive jet data and found to be of the order 
of $10^{-4}-10^{-3}$ per jet 
for $15 < \et < 100$\gev.
These misidentification rates are used to weight the 
simulated events of the background processes, resulting in a total 
background yield estimated to be less than 0.01 event.

The acceptance for standard model 
$\ppbar \to Z/\gamma^*\,Z/\gamma^*\to\ell^+ \ell^- \ell^+ \ell^-$ 
is determined using the leading-order {\sc pythia} generator 
and found to be $0.17\pm 0.02$.
The uncertainty has contributions from higher-order generator effects, 
lepton identification, and trigger efficiency uncertainty.
In order to estimate the 
uncertainty arising from higher-order generator effects, the {\sc mc@nlo} generator \cite{mcnlo} 
is used, interfaced to {\sc herwig} \cite{herwig}
to provide parton showering and hadronization.
The corresponding relative uncertainty on the acceptance is 
estimated to be 2.7\%.
Lepton identification efficiencies are measured in the data using 
candidate \zll\ events with uncertainties at the level of 1\%.  
We also account for a small drop in lepton identification efficiency with 
time and assign a 2\% relative uncertainty per lepton for residual 
run-dependent effects.
We assume no correlation between the uncertainties on electron and muon 
reconstruction, and full correlation between the uncertainties for 
leptons of the same flavor.
The trigger efficiency 
per four-lepton event is close to unity, with a systematic uncertainty 
of less than 0.5\%. 

Given the branching fraction for $\zll=(3.366\pm 0.002)\%$ \cite{pdg}, 
the branching fraction for two \z\ bosons to decay to electrons or muons 
is $4.52\times 10^{-3}$.  The scale factor to take into account differences 
in triggering, reconstruction and identification efficiencies between 
data and simulation is $0.80\pm 0.08$, and the integrated luminosity is
$5.91\pm 0.35$\,fb$^{-1}$.
Experimentally, we observe $\ppbar \to Z/\gamma^*\,Z/\gamma^*\to\ell^+ \ell^- \ell^+ \ell^-$, and 
to compare our measurement with the theoretical prediction of
$\ppbar \to ZZ$, calculated in a narrow pole approximation \cite{NLO_CROSS_SECTIONS}, 
we account for $Z/\gamma^*$ interference.   The interference in the region 
$76 < \mll < 106\gevcsq$ increases the acceptance by a factor of 1.03.
From simulation, the fraction 
of \zz\ events that falls outside the region $76 < \mll < 106\gevcsq$ is 0.07 
and is also corrected for.
The eight observed events therefore result in a cross section: 
$$
      \sigma(\ppbar \to ZZ) ~=~ 
          2.3 ~ ^{+0.9}_{-0.8} ~\stat ~\pm~ 0.2 ~\syst\pb
$$
where the statistical uncertainty is the 68\% confidence interval 
given by the method of Feldman and Cousins \cite{feldmancousins}.
The value is consistent with the theoretical 
prediction  $1.4\pm 0.1$\pb\ \cite{NLO_CROSS_SECTIONS}.
A more precise measurement of the \zz\ cross section, 
which combines four-lepton and leptons plus \met\ channels, 
is reported elsewhere \cite{matteo}.

\begin{figure}[h!]
  \begin{center}
    \includegraphics[width=0.37\textwidth, clip=true, viewport=0.1in 0.2in 7.8in 6.5in] 
    {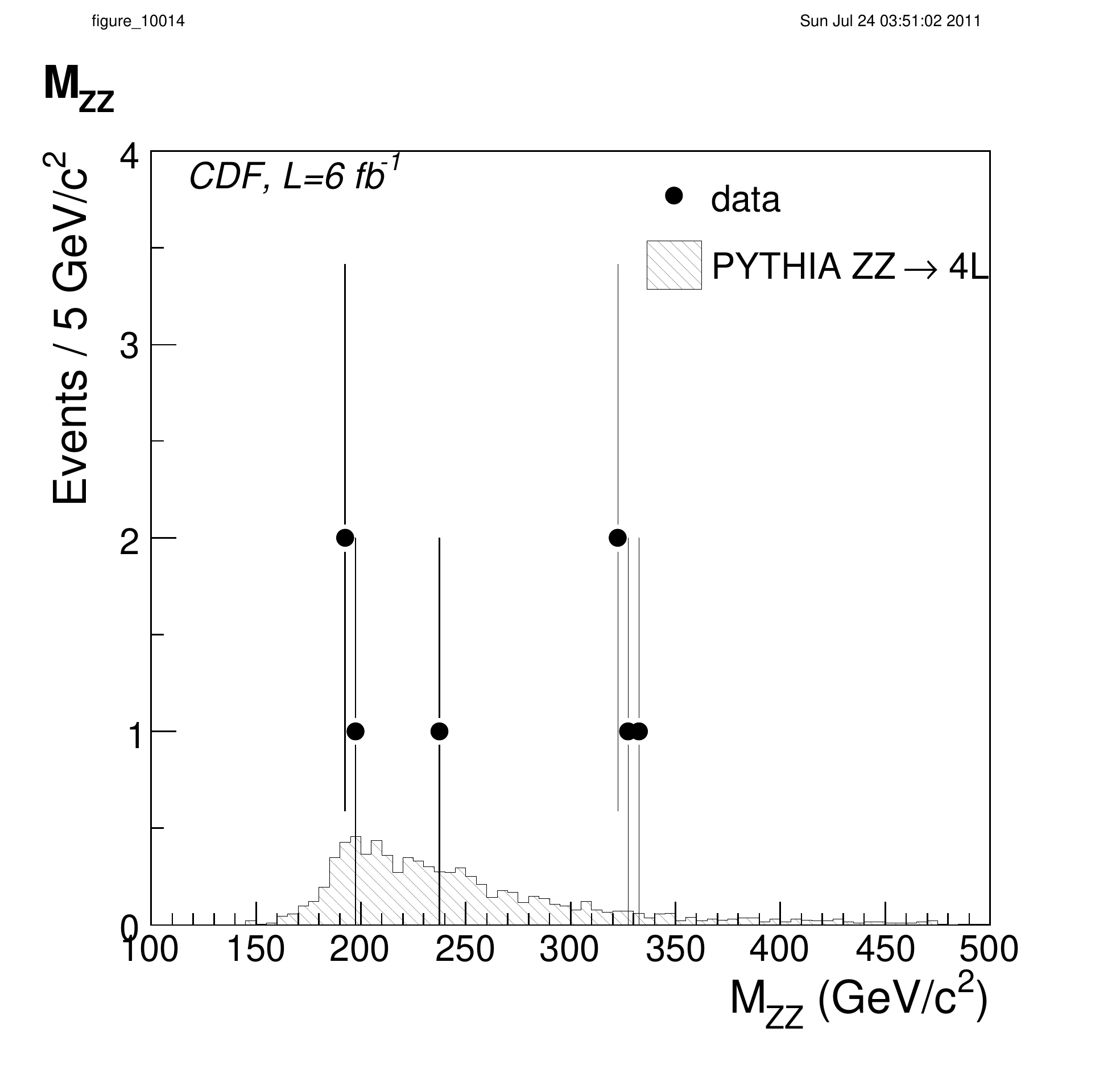}
    \caption[]{ 
      \mzz\ for eight \zzllll\ candidates ({\sc pythia} normalized 
      to the standard model prediction of 5.5 events).
    }
    \label{fig:mzz}
  \end{center}
\end{figure}

Examining the properties of the eight \zz\ candidate events we find 
an excess of events over standard model expectations at high 
invariant mass, \mzz. 
The invariant masses of four events are clustered
with mean 327\gevcsq, 
as shown in Fig.\,\ref{fig:mzz}.
All four candidates,
one \eeee, one \eemm, and two \mmmm, have values 
of \mzz\ within 7\gevcsq\ of the mean.  In the four-lepton channel
the detector resolution in \mzz, 
$\sigma(\mzz)$, is 5 to 6\gevcsq, so within detector resolution 
the masses of all four events are consistent 
with a potential new resonance.

To study the possibility that these events are due to a decay of a heavy 
resonance, we split the eight candidate events into low- and high-mass 
samples and compare the properties of the events in the two samples.
The high-mass region is defined by an
a posteriori choice $\mzz>300\gevcsq$, which is 
$\sim 5 \sigma(M_{ZZ})$ below the observed clustering of events;
less than 25\% of the expected standard model \mzz\ distribution 
lies above this cutoff.

\begin{figure}[h!]
  \begin{center}
    \includegraphics[width=0.5\textwidth, clip=true, viewport=0.1in 0.in 7.7in 3.3in] 
    {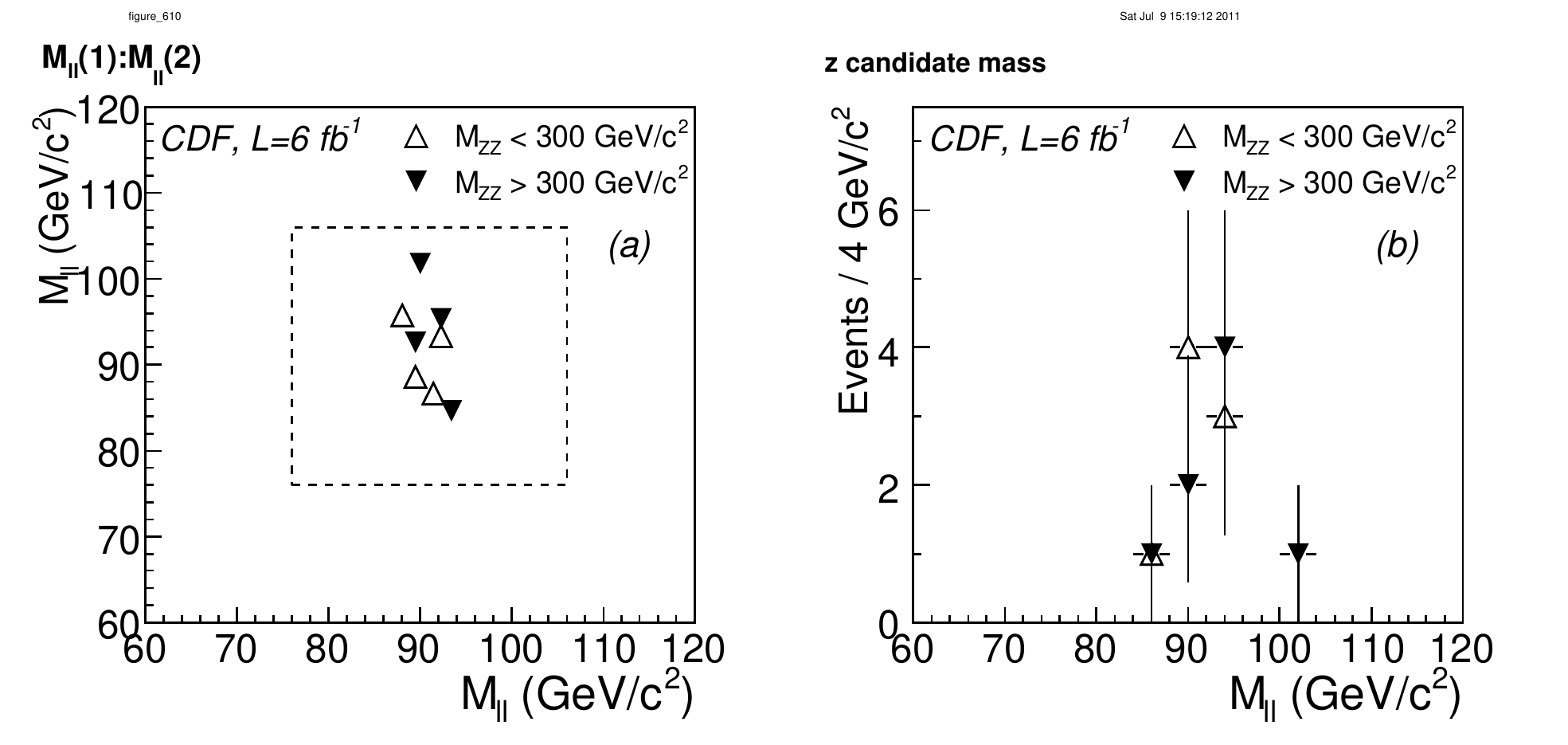}
    \caption[]{ 
      Invariant masses of dilepton pairs for eight \zz\ candidate events:  
      (a) \mll(1) versus \mll(2), with selected mass region outlined; and (b) \mll\ for all \z\ boson candidates.
    }
    \label{fig:mz}
  \end{center}
\end{figure}

\begin{figure}[h!]
  \begin{center}
    \includegraphics[width=0.5\textwidth, clip=true, viewport=0.1in 0.in 7.7in 3.3in] 
    {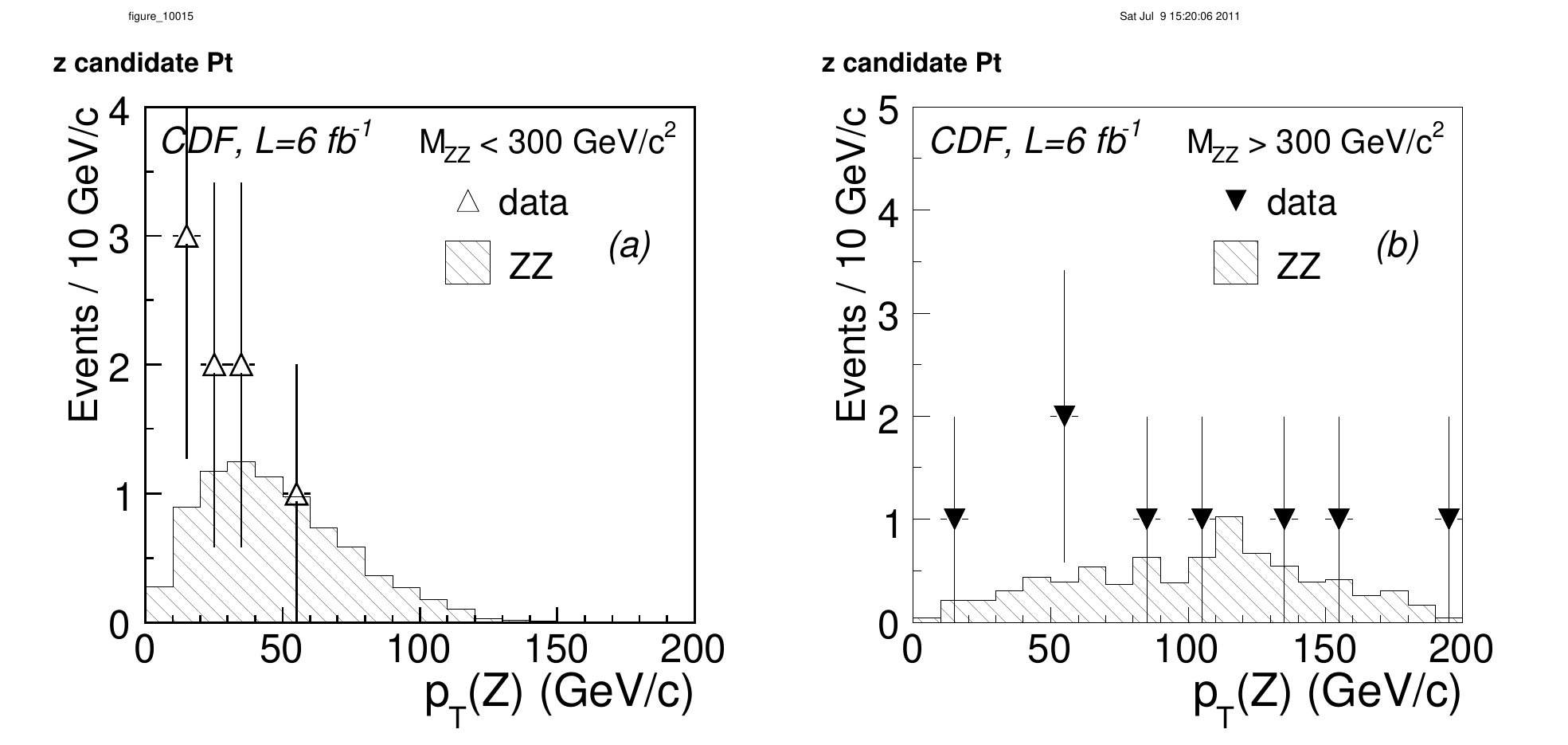}
    \caption[]{ 
      $p_T(\z)$ for \z\ boson candidates in (a) low-mass four-lepton candidate 
      events and (b) high-mass events ({\sc pythia} prediction 
      normalized to four events in each plot).
    }
    \label{fig:ptz}
  \end{center}
\end{figure}

The masses of the \z\ boson candidates for all events are
shown in Fig.~\ref{fig:mz}, which demonstrates that the 
resolution in $M_{\ell\ell}$ is consistent in the high-mass and 
low-mass events.
Lepton identification variables are consistent with expectation 
for all the observed events.
Most kinematic distributions for the \zzllll\ candidates are 
in agreement with standard model 
expectations; as one example, the $p_T$ distributions of the 16 \z\ boson candidates
are shown in Fig.~\ref{fig:ptz}.

\begin{figure}[h!]
  \begin{center}
    \includegraphics[width=0.5\textwidth, clip=true, viewport=0.1in 0.in 7.7in 3.3in] 
    {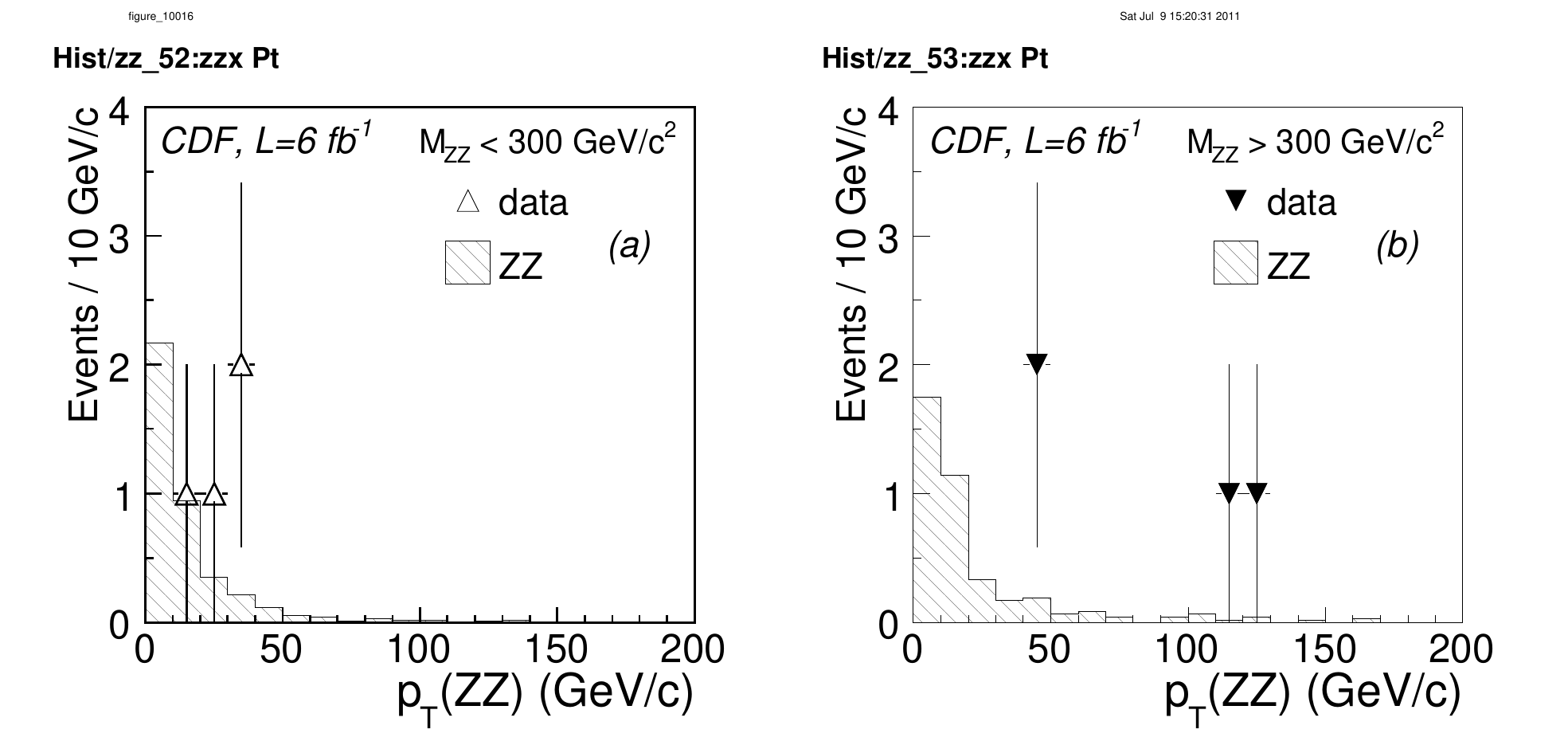}
    \caption[]{ 
	$p_T(\zz)$ for (a) low-mass four-lepton candidate events  
        and (b) high-mass events ({\sc pythia} prediction normalized 
        to four events in each plot).
    }
    \label{fig:ptzz}
  \end{center}
\end{figure}

\begin{figure}[h!]
  \begin{center}
    \includegraphics[width=0.5\textwidth, clip=true, viewport=0.1in 0.in 7.7in 3.3in] 
    {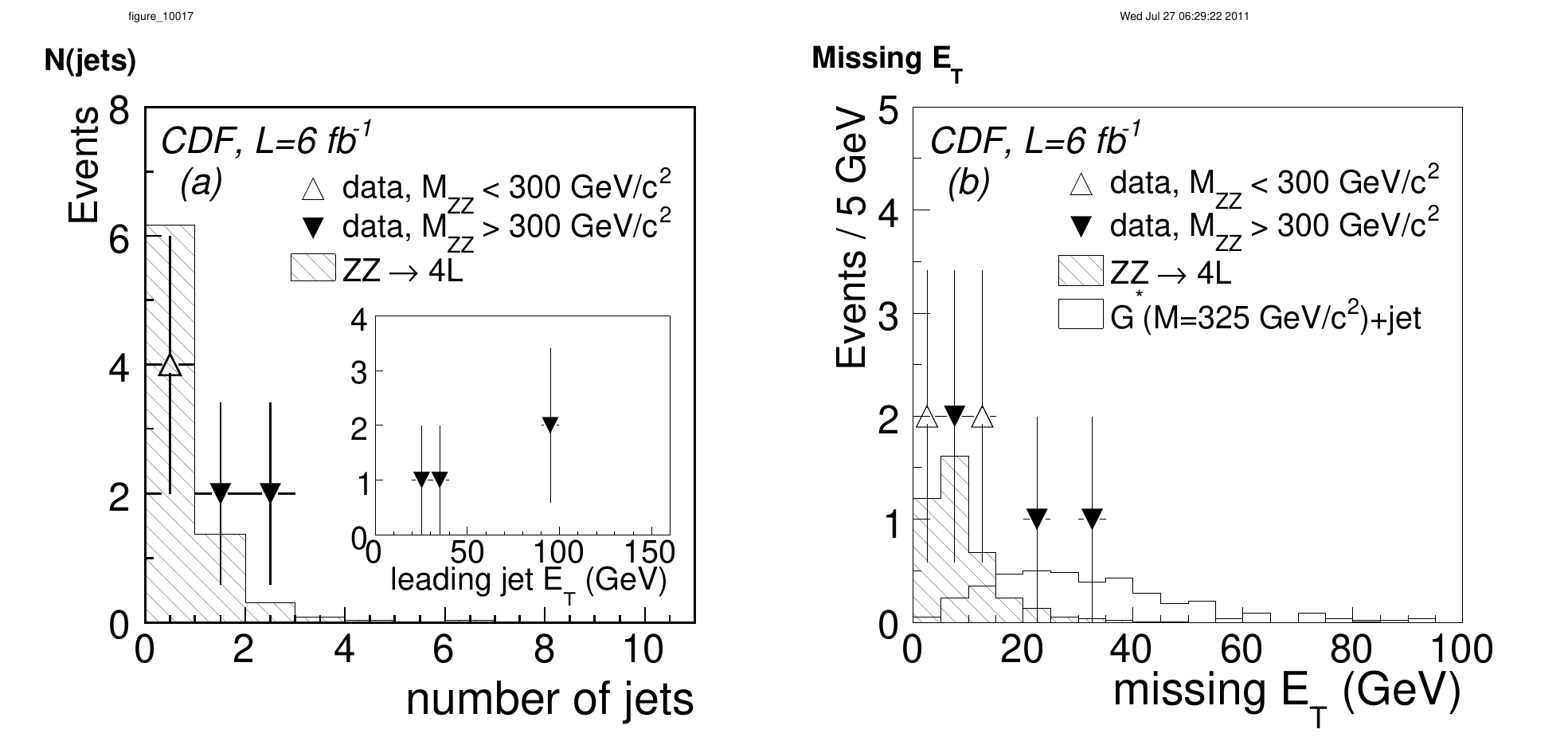}
    \caption[]{ 
	(a) Number of jets and (inset) \et\ of the most energetic jet; 
        and (b) \met\ for four-lepton candidate events. \met\ distribution for G*(+jet) 
        process is normalized to 4 events.
    }
    \label{fig:ptjetmet}
  \end{center}
\end{figure}

However, for the high-mass events, the $p_T$ distribution of the 
four-lepton system is rather different from the standard model 
expectation, as shown in Fig.~\ref{fig:ptzz}. 
The \zz\ system in the high-mass events is seen to be boosted and, 
as shown in Fig.~\ref{fig:ptjetmet}, is recoiling against 
one or more jets. 
None of the four low-mass events has a reconstructed jet with \et\ above 20\,GeV.

We check whether there is any indication of misreconstruction in 
these events.  In \zzllll\ events, where there is no real \met,
large measured \met\ could indicate misreconstruction.
However the presence of jets broadens the detector \met\ resolution 
and needs to be taken into account.  
To this end we exploit two physics models.
The first model is RS graviton production through
gluon-gluon fusion (the `$s$-channel signal model') \cite{lisafitzpatrick}.
In order to investigate effects of the production mechanism and 
in the absence of a
particular model that would predict the production of a boosted \zz\ resonance,
we take as an alternative signal model the production of a Kaluza-Klein
excitation of a graviton, $G^*$, of \mgstar\ 
recoiling against a
parton of $\et\geq 100$\gev (referred to as the `boosted signal model').
In both cases the {\sc herwig} event generator is used with the full CDF
detector simulation.
In the four-lepton decay channel, neither of these models generates real \met. 
Fig.~\ref{fig:ptjetmet}(b) thus 
demonstrates that the observed \met\ in the 
high-mass events is consistent 
with resolution effects arising from the jets.

Overall, we conclude that the observed events are well-measured and that,
within the detector resolution, the kinematic parameters of the \z\ candidates
are reconstructed correctly. The event properties are given in
Table~\ref{table:4l_candidate_events_01}.

\begin{table*}[tb]
\caption{  Properties of the four-lepton candidate events, in the order in which they were recorded.
   \label{table:4l_candidate_events_01} 
} 
\small
  \begin{center}
    \begin{tabular}{cccccccc}
      \hline                                                             
      \hline                                                             
leptons&$M_{Z_1}, p_T(Z_1)$&$M_{Z_2}, p_T(Z_2)$ & \mzz &$p_T(\zz)$& \met &$N_{jets}$& Jet \et   \\
       & (\gevcsq), (\gevc) & (\gevcsq), (\gevc) & (\gevcsq) & (\gevc) & (\gev) & &  (\gev) \\
\hline                                                                       
 \eeee  &  93.3, 18.2     &   92.9, 17.4     &   196.6    &  35    &  14  &    0    &            \\
 \mmmm &  85.9, 101.9    &   92.1, 54.8     &   321.1     &  47.4  &   8.4&    1    & 36.7       \\
 \eemm &  92.0, 156.0    &   89.9, 139.7    &   324.7     & 126.8  &  31  &    2    & 97.4, 40.0 \\
 \eeee  & 101.3, 57.8     &   91.6, 13.2     &   334.4    &  44.7  &  9.9 &    1    & 22.7       \\
 \eemm &  87.9, 17.7     &   91.8, 29.8     &   191.8     &  31    &  10.5&    0    &            \\
 \mmmm &  95.9, 197.9    &   92.0, 87.2     &   329.0     & 110.9  &  23.3&    2    & 97.2, 24.7 \\
 \eemm &  95.2, 36.7     &   89.7, 38.8     &   237.5     &  10.2  &   1.2&    0    &            \\
 \mmmm &  88.4, 51.0     &   89.8, 26.6     &   194.1     &  25.9  &   3.3&    0    &            \\
%
%
\hline   
\hline   
\end{tabular}     
\end{center}  
\end{table*}

To quantify consistency between the data and the standard model, 
we compute the probability for the observed \mllll\ distribution 
to be due to a statistical fluctuation of the standard model expectation. 
Eight-event pseudoexperiments are drawn from the standard model \mzz\ 
distribution, and a test statistic is computed for each pseudoexperiment.

Two tests are performed.
First, the Kolmogorov-Smirnov (KS) distance is taken as the test statistic, 
with the intention of testing for goodness-of-fit in a general way.
The fraction of pseudoexperiments that has KS distance 
greater than that of the observed data distribution determines the 
computed $p$-value, which is found to be 0.14.

Second, a more powerful test statistic for a resonance 
search is used: the ratio of likelihoods of two hypotheses.
The background hypothesis is provided by the standard model distribution
in \mzz, $M^{SM}_{ZZ}$, and the signal hypothesis adds to it a resonance 
represented by a Gaussian peak: $f\cdot M^{SM}_{ZZ} + (1-f)\cdot G(M,w)$.
For a given mass $M$, the resonance width $w$ is defined by the detector resolution 
at this mass.
The resonance parameters are defined from fitting the pseudoexperiment 
distribution in \mzz.
The likelihood ratio for the data is computed using the same procedure.
The fraction of pseudoexperiments that has likelihood ratio 
L$_{\rm SM}$/L$_{\rm SM+G}$
lower than that of the observed data distribution determines the 
computed $p$-value and is found to be $(1-2)\times 10^{-3}$,
where the range comes from shape differences of the 
{\sc pythia} and {\sc mc@nlo+herwig} event generators.

In the absence of a physics model that would predict the observed 
\pt(\zz) distribution, we quantify consistency between 
the data and the standard model by computing the probability for 
eight events sampled from the 
standard model \pt(\zz) distribution to have KS distance greater than that 
observed in the data.
The probability 
for the data to represent the standard model distribution is
$(1-2)\times 10^{-4}$.%


\section{ \zzllnn\ channel}
\label{zzllnn}

The four-lepton events observed above 300\gevcsq\ appear somewhat anomalous.
If these events were to be due to a new \zz\ resonance, it would also 
be detectable in the other \zz\ decay modes, $\ell\ell\nu\nu$ and \lljj. 
\z\ bosons coming from the decay of such a heavy particle would 
be boosted, so
events with one of the \z\ bosons decaying into neutrinos would have large \met. 
For each lepton flavor, the branching ratio to neutrinos is about twice 
that of charged leptons.  With all three neutrino flavors included, 
and only one \z\ boson to be reconstructed, the expected event yield 
is around ten times higher than in the four-lepton channel, and the
sensitivity to new physics at $\mzz=325$\gevcsq\ is several
times better than in the four-lepton channel.

Optimising sensitivity for a resonance of mass $\mzz\sim 325\gevcsq$,
we define the search region to be $\met>100$\gev. 
The standard model expectation for events with a \zll\ candidate 
and such high \met\ is around 25 events, as given in 
Table~\ref{table:llvv}.
\zee\ and \zmumu\ candidates are selected 
according to the requirements described for the \zzllll\ channel.
Owing to the extra acceptance, we did not reprocess the \llmet\ data.

We validate the background model using events with a 
reconstructed \z\ boson and $\met<100$\gev.
Irreducible background contributions to a search for new physics
in this channel come
from standard model diboson production processes \ww, \wz, and \zz, as 
well as from \ttbar\ production.
Other non-negligible background contributions come from \z+jets events that 
have large \met\ due to jet mismeasurement; 
from $W$+jets events where one of the jets is misreconstructed as 
a lepton and forms a \z\ boson candidate with the charged 
lepton from the decay of the $W$ boson; 
and, in the $ee+\met$ channel, from 
$W\gamma$ production with the photon misreconstructed as an electron.

\begin{figure}[h!]
  \begin{center}
    \includegraphics[width=.49\textwidth, clip=true, viewport=0.1in 0.in 7.7in 3.4in] 
    {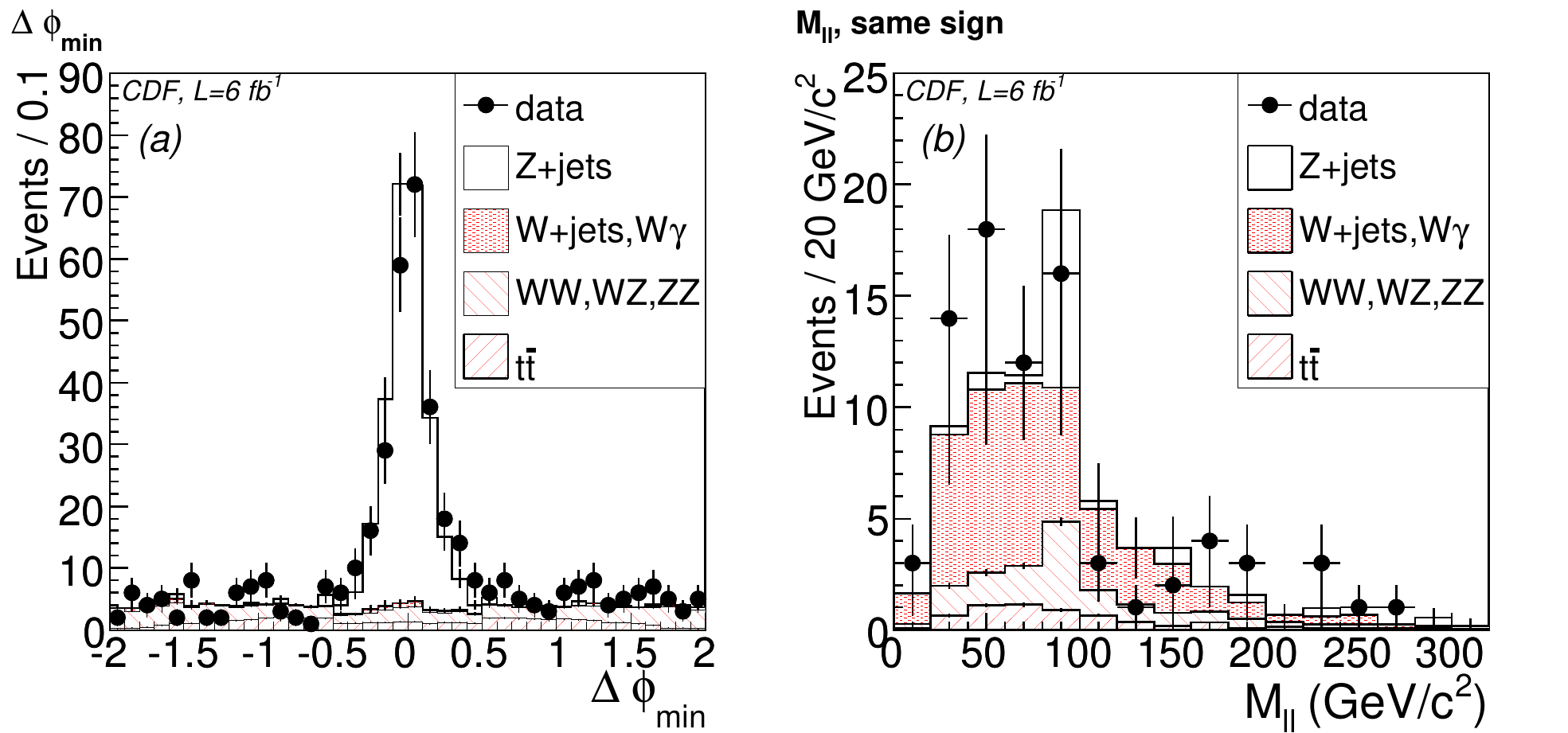}
    \caption[]{
      (a) $\Delta\phi_{\rm min}$ as used for \z+jets normalization, 
      and (b) \mll\ for same-sign dielectron pairs with large \met\ 
      used to validate the $W$+jets background estimation.}
    \label{fig:llvv_dphi}
  \end{center}
\end{figure}

Irreducible backgrounds are estimated using the {\sc pythia} 
generator and the full CDF detector simulation, 
normalized to NLO cross sections \cite{NLO_CROSS_SECTIONS}.
The \z+jets contribution is also estimated using {\sc pythia} 
simulation and is normalized using a subset of the $\met<100\gev$ data.
As \z+jets events have high \met\ only through misreconstruction, 
the normalization is carried out on events having $50<\met<100$\gev\
that also have a small angle $\Delta\phi_{\rm min}$ between the \met\ 
and the closest jet, or lepton, reconstructed 
in the event: $|\Delta\phi_{\rm min}|<0.5$.
The $|\Delta\phi_{\rm min}|$ distribution is shown in Fig.~\ref{fig:llvv_dphi}(a). 
It is verified that this procedure is not sensitive to the \met\ range used.

The background contribution from the $W$+jets process is estimated 
from a data sample where events contain an identified lepton 
and an additional jet.  
These events are weighted by jet-to-lepton 
misidentification rates as described in Section \ref{section:fakerates}
to estimate the total yield.
Owing to differences in jet-to-lepton fake rates between electrons and muons,
the $W$+jets contribution is found to be negligible in the 
$\mu\mu+\met$ channel, but non-negligible in the $ee+\met$ channel.

Photon conversions are the primary source of jets being misidentified 
as electrons, and so $W$+jets events 
result in approximately equal numbers of same-charged and oppositely-charged
candidate events. 
The estimate is therefore validated against the sample of events
that have two lepton candidates of the same charge and 
$50<\met<100$\gev.
Fig.~\ref{fig:llvv_dphi}(b) shows that this selection is dominated by $W$+jets.
The estimate is also cross-checked by applying the same misidentification 
rates to \wenu\ simulation normalized to the NLO production cross section. 
The two methods give results consistent within 10\%.

\begin{figure}[h!]
  \begin{center}
    \includegraphics[width=.35\textwidth, clip=true, viewport=0.in 0.in 7.7in 6.9in] 
       {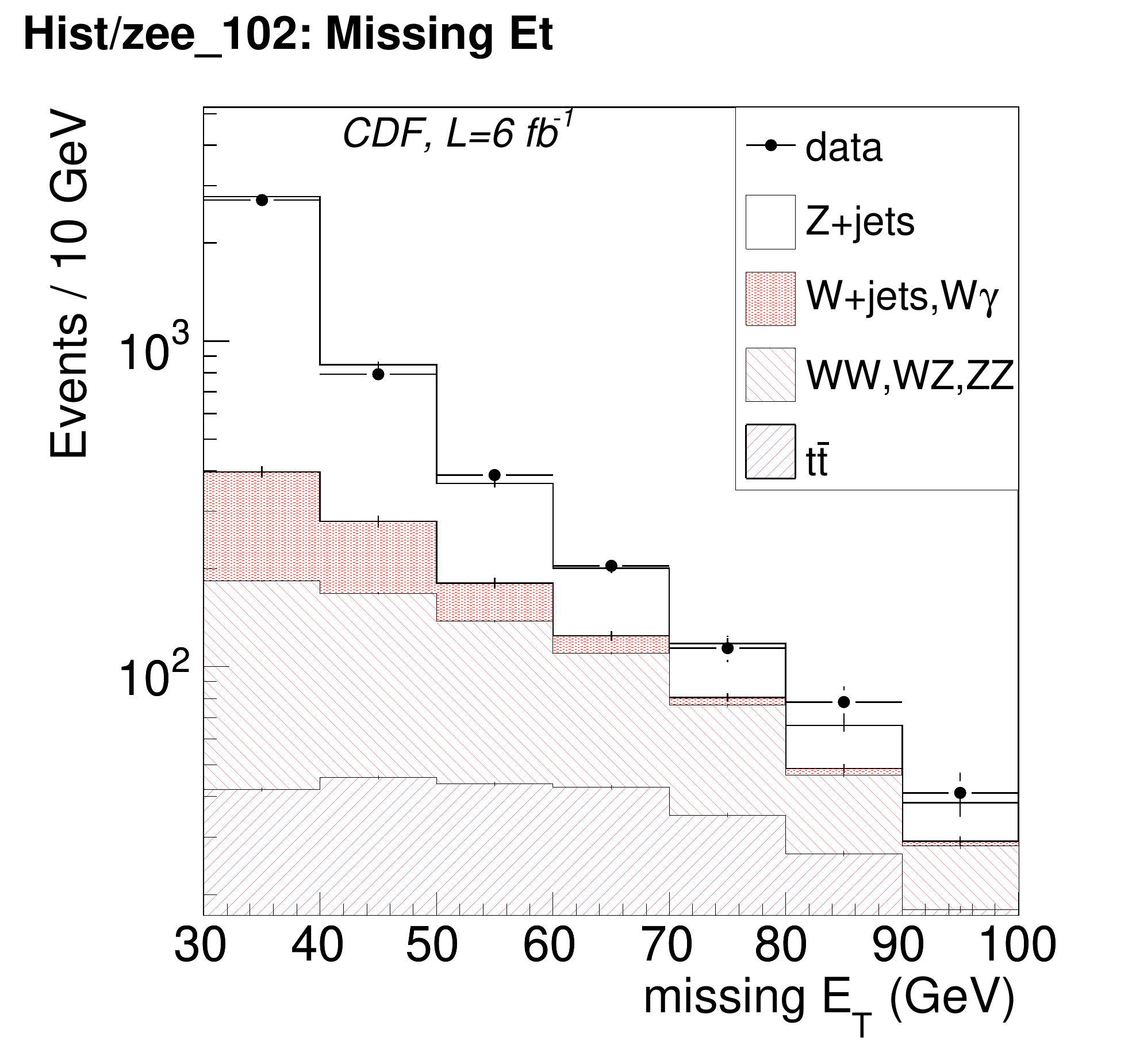}
       \caption[]{
         \met\ distribution for events with opposite sign lepton pairs ($ee+\mu\mu)$.
         The contribution of $Z$+jets events is normalized in the region 50 \lt\ \met\ 
         \lt 100\gev\ using events with low $|\Delta\phi_{\rm min}|$.
       }.
       
    \label{fig:llvv_met}
  \end{center}
\end{figure}

The overall modeling of the sample composition is demonstrated by the 
\met\ spectrum shown in Fig.~\ref{fig:llvv_met}.
The largest relative uncertainty in this channel comes from the 
\z+jets normalization, and is 10\% and 13\% in the electron and 
muon channels respectively.
Other uncertainties come from lepton identification (2\%), acceptance (\lt 1\%), 
cross sections of diboson and top-quark production (5\% and 10\%), 
and the fake lepton background (20\%).   
The total background uncertainty is 13\%.

To search for a high-mass resonance we examine events with 
$\met>100$\gev. 
Event yields are given in Table~\ref{table:llvv}.
In electron and muon channels combined we expect 
26 events from standard model processes, and observe 27.
Four four-lepton events around $\mzz=325$\gevcsq\ coming from 
the decay 
of a new state would imply a production cross section times 
branching ratio to \zz\ close to 1\pb, and for that cross section, 
the $s$-channel $G^*$ signal model predicts around 35 additional events.

\begin{figure}[h]
  \begin{center}
    \includegraphics[width=.49\textwidth, clip=true, viewport=0.1in 0.in 7.7in 3.4in] 
    {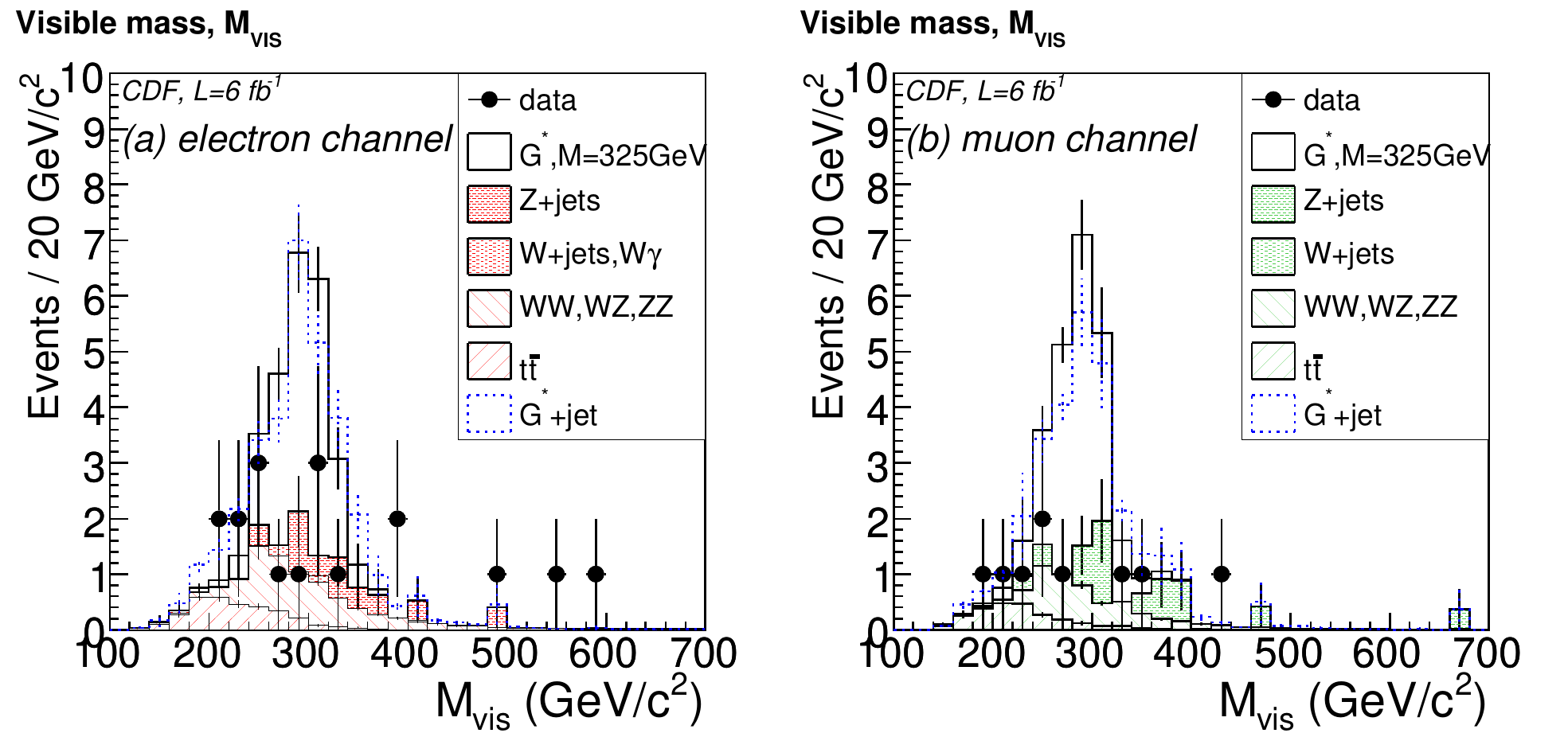}
    \caption[]{
      \mvis\ for (a) the electron and (b) muon channels.  
      The expected contribution from a graviton of \mgstar\ and 
      cross section times branching ratio to \zz\ of 1\pb\ 
      is shown together with the expected contribution of boosted $G^*$, 
      produced in association with a jet.
      The high values of \mvis\ of three events in the electron channel are
      understood as originating from fluctuations of the jet energy losses 
      in events with high jet activity.
    }
    \label{fig:llvv_sig_summary}
  \end{center}
\end{figure}

As the second \z\ boson in this channel decays into neutrinos,
the invariant mass of the \z\ pair  cannot be fully reconstructed.
The closest approximation is the `visible mass' \mvis, 
defined as the invariant mass of the sum of the two charged lepton
four-momenta 
and the four-vector representing the \met, $(\metx, \mety, 0, |\met|)$.
Fig.~\ref{fig:llvv_sig_summary} shows the \mvis\ distribution 
in the signal region, $\met>100$\gev, with the expected distribution 
for an RS graviton of mass $M_{G*}=325$\gevcsq\ and cross 
section times branching ratio of 1\pb\ overlaid.
In this channel we find little difference in expected 
distributions or yields between the two signal models, 
confirming that the analysis is not strongly dependent on the 
detail of the models.
Neither the event counts of Table~\ref{table:llvv}, nor the distributions 
of Figure~\ref{fig:llvv_sig_summary}, show any evidence for a resonance 
decaying into \zz.

\begin{table}[h]
\caption{  Expected and observed event yields in the \llmet\ channel.
   \label{table:llvv} 
}
\centering{
\resizebox{   .48\textwidth}{!} {
\begin{tabular}{ccc}
\hline
\hline
Source   & electron channel & muon channel \\
  \hline
\zz      & 1.8    &  1.3    \\
\wz	 & 3.6    &  2.8    \\
\ww      & 0.9    &  0.5    \\
\ttbar	 & 3.2    &  2.4    \\
$W$+jets & 0.1    &  0.3    \\     
$Z$+jets & 4.0    &  5.1    \\
  \hline 	
  \hline 	
Total standard model & $13.6\pm 1.8$  &  $12.4\pm 1.6$  \\
  \hline 
Data     &  18    &  9  \\
  \hline 
  \hline 
Expected $s$-channel signal,               & \\
$M_G=325$\gevcsq\ and $\sigma$=1\,pb   & $17\pm 1$ &  $18\pm 1$ \\
  \hline 
Expected boosted signal,               & \\
$M_G=325$\gevcsq\ and $\sigma$=1\,pb   & $20\pm 1$ &  $17\pm 1$ \\
  \hline 
  \hline 

\end{tabular}
}}
\end{table}


\section{ \zzlljj\ channel}
\label{zzlljj}

The decay of a heavy particle into two \z\ bosons 
where one of the \z\ bosons decays into charged leptons and 
the other to jets has the 
advantage of being fully reconstructible, and the event yield in 
the \lljj\ channel is expected to be around twenty times 
higher than in the four-lepton channel.

\zee\ and \zmumu\ candidates are selected according 
to the requirements described for the \zzllll\ channel, 
and a further requirement is made of at least two 
reconstructed jets having corrected \et \gt 25\gev.  
To reconstruct the second \z\ boson candidate, all pairs of jets 
are considered and if there is a pair with invariant 
mass between 70 and 110\gevcsq\ it is accepted.
This inclusive selection, with the additional requirement 
of the invariant mass of the two \z\ candidates 
being less than $300$\gevcsq, defines a control region.

This channel is dominated by \z+jets events.
Other standard model sources, small compared with \z+jets, are
\wz\ and \zz\ production, and \ttbar\ production.
The contributions from \ww\ and $W$+jets events are negligible.  

\begin{figure}
  \begin{center}
    \includegraphics[width=.5\textwidth, clip=true, viewport=0.in .96in 7.8in 4.5in] 
       {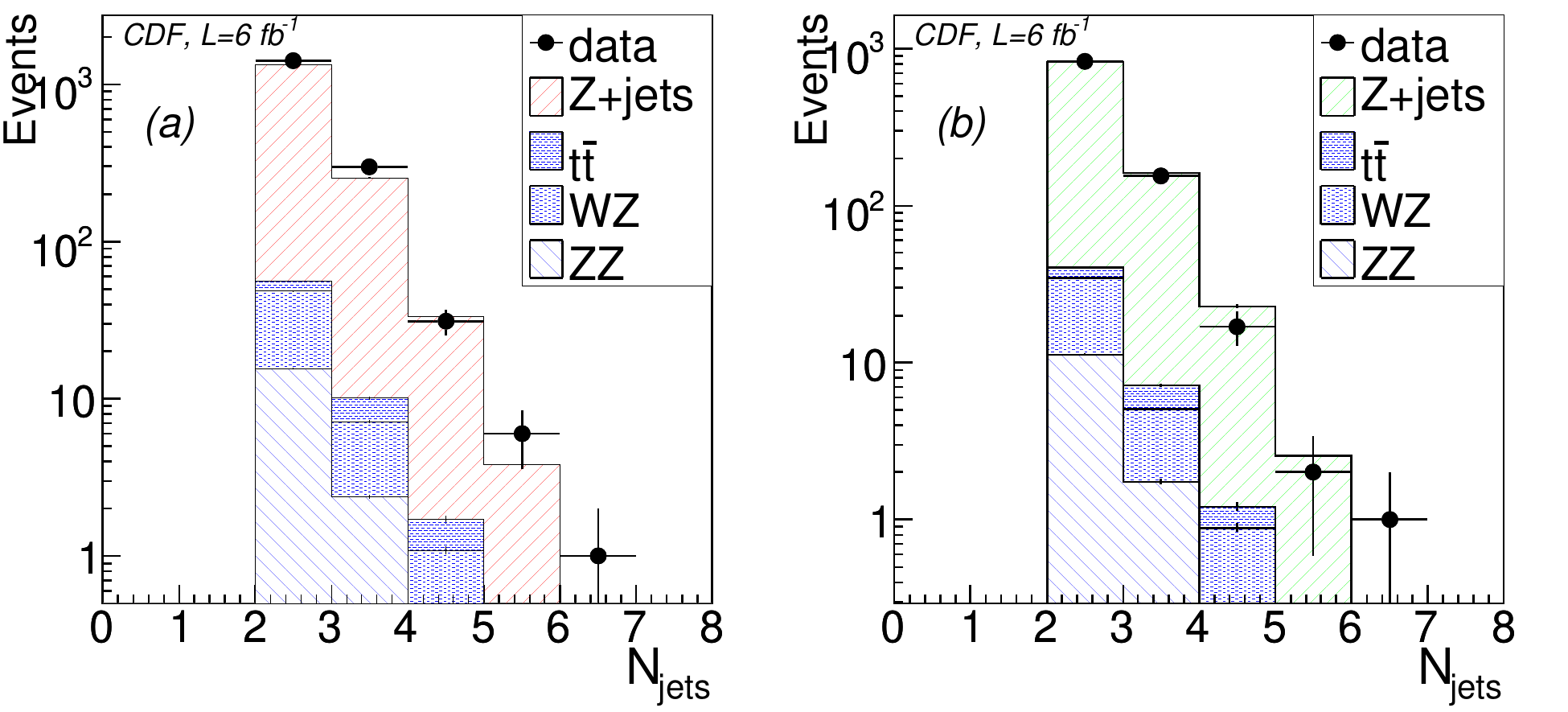}
       \caption[]{Number of jets in (a) $\zee+\geq 2$\,jets and (b) $\zmumu\geq 2$\,jets events in the control region $\mlljj <300\gevcsq$.}
    \label{fig:njets}
  \end{center}
\end{figure}

Diboson and \ttbar\ event yields are estimated using 
{\sc pythia} Monte Carlo normalized to NLO cross sections.
\z+jets events are modeled using the generator 
{\sc alpgen} \cite{alpgen} interfaced with {\sc pythia} for parton 
showering and hadronization, and the normalization of the \z+jets contribution 
is obtained by fitting to the total data yield 
in the control region.
The detector acceptance is different for \zee\ and \zmumu and
so the \z+jets normalization factors for the two channels are not 
expected to be identical.  The difference between them is 
indicative of the systematic uncertainty, leading to a 
total background uncertainty of 10\%.
The jet multiplicity distributions in the control region, 
shown in Fig.~\ref{fig:njets}, demonstrate the good background modeling.

In the \lljj\ final state we improve the resolution in 
the reconstructed \mzz\ by varying jet four-momenta
within their uncertainties and constraining the reconstructed 
invariant masses \mjj\ to 
the mass of the \z\ boson, $M_{Z}$.
The resolution in $M_{Z}$ for \zjj\ is $15\gevcsq$, which 
is much larger than the intrinsic width of the \z\ boson.
In the \lljj\ channel the constraining procedure therefore improves the mass 
resolution of the \zz\ candidates, to $12\gevcsq$ for \mgstar.
As the detector resolution for $M_{Z}$ in \zll\ is comparable with the 
intrinsic width of the \z\ boson, applying the mass-constraining procedure 
to the leptons 
has very little effect on the \mzz\ resolution 
and is used only as a cross-check.
Throughout this paper 
\mlljj\ refers to the constrained four-object invariant mass.

\begin{figure}[h]
  \begin{center}
    \includegraphics[width=.32\textwidth, clip=true, viewport=0.in .96in 3.9in 4.8in] 
       {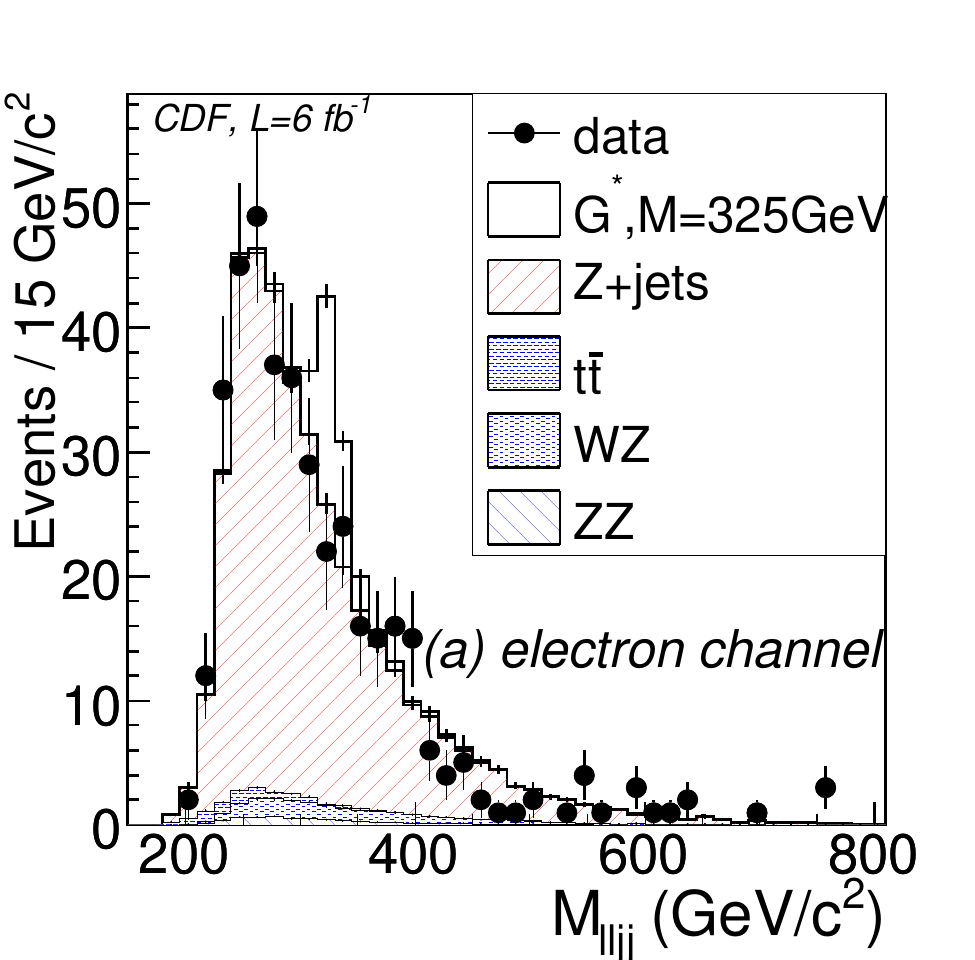}
    \includegraphics[width=.32\textwidth, clip=true, viewport=3.9in .96in 7.8in 4.8in] 
       {\plots/summary_1_sig}
       \caption[]{\mlljj\ for the (a) electron and (b) muon channels, showing the expected contribution from a graviton of \mgstar\ and cross section times branching ratio to \zz\ of 1\,pb.}
    \label{fig:lljj_sig_summary}
  \end{center}
\end{figure}

To search for a high-mass resonance we examine the complete 
\mlljj\ spectrum. 
\z\ bosons coming from the decay of a heavy particle would 
be boosted, and optimization studies result in 
requiring the most energetic
jet in the \zjj\ candidate to have $\Et>50$\gev\ and the 
\pt\ of either the \zjj\ or \zll\ candidate to
be greater than 75\gevc.
Observed event yields are given in Table~\ref{table:lljj} and are consistent 
with standard model expectations.
A resonance of \mgstar\ and cross section times branching 
ratio to \zz\ of 1\pb\ would be expected 
to yield around 30 events in the muon channel and 40 events in the 
electron channel, and
as the \zzlljj\ final state is fully reconstructed, they would 
appear as a narrow peak in \mlljj.
Fig.~\ref{fig:lljj_sig_summary} shows the \mlljj\ distribution 
for the \eejj\ and \mmjj\ channels, with the standard model and 
additional \zz\ resonance model predictions.

\begin{table}[h]
\caption{  Expected and observed event yields in the \lljj\ channel.
   \label{table:lljj} 
}
\centering{
\resizebox{   .48\textwidth}{!} {
\begin{tabular}{ccc}
\hline
\hline
Source   & electron channel & muon channel \\
  \hline
\zz         &     6  &    5    \\
\wz	   &    17  &   12    \\
\ttbar	   &     7  &    5    \\
$Z$+jets    &   395  &  244    \\
  \hline 	
  \hline 	
Total standard model	   &   424$\pm$40  &  266$\pm$24  \\
  \hline 
Data       &   392  &  253  \\
  \hline 
  \hline 
Expected signal,               & \\
$M_G=325$\gevcsq\ and $\sigma$=1\,pb   &41$\pm$1 & 32$\pm$1 \\
  \hline 
  \hline 

\end{tabular}
}}
\end{table}

\begin{figure}[h]
  \begin{center}
    \includegraphics[width=.5\textwidth, clip=true, viewport=0.in 0.96in 7.8in 4.8in] 
       {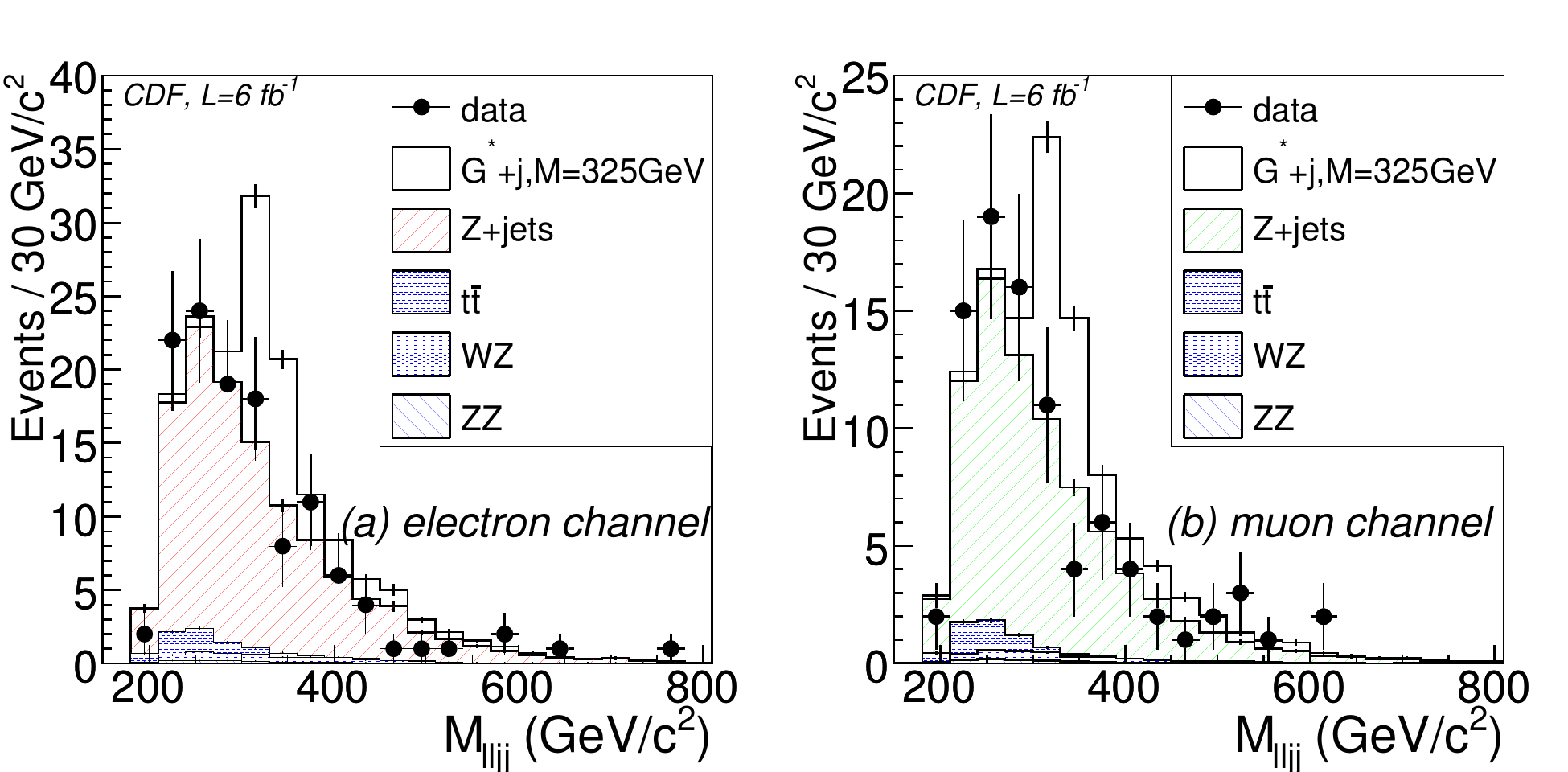}
       \caption[]{\mlljj\ for the (a) electron and (b) muon channels for $\pt(\zz)>40$\gevc, showing the expected contribution from a boosted graviton of \mgstar\ and cross section times branching ratio to \zz\ of 1\,pb.}
    \label{fig:lljj_dd_summary}
  \end{center}
\end{figure}

Studies of systematic effects resulting from the generator 
$Q^2$ scale choice and from the jet energy scale uncertainty show that 
they do not affect the expected shapes of the \mlljj\ distributions.
We investigate potential effects of the production mechanism 
using the alternative boosted $G^*$ signal model.  
Motivated by the anomalous \pt(\zz) distribution shown by the 
events in the four-lepton channel, 
the signal selection is modified to require $\pt(\lljj)>40$\gevc, 
which further suppresses standard model background,
The resulting 
\mlljj\ distribution and boosted $G^*$ prediction 
is shown in Fig.~\ref{fig:lljj_dd_summary}.
As with the \llmet\ 
channel there are no statistically significant 
differences from the standard model expectation.

\section{Limits}
\label{section:limits}

To quantify results of the search we compute expected 
and observed limits on the production cross section times 
branching ratio $\sigma(\ppbar \to G^* \to ZZ)$.

The expected sensitivity is determined with a 
Bayesian technique \cite{joel}, using CL$_S$ likelihood test statistics \cite{junk} 
to perform a binned maximum-likelihood fit over the 
\mzz, \mvis, and \mlljj\ distributions in the four-lepton, \llmet, and 
\lljj\ channels respectively. 
The background hypothesis is provided by the standard model expectation
as described in Sections \ref{zz4l}-\ref{zzlljj}.
Background-only pseudo-experiments are drawn from Monte Carlo 
simulation.
In the fit, the background templates can fluctuate within their uncertainties.
A test statistic is formed from the difference in the likelihoods 
between the background-only model and the signal-plus-background 
model at the best fit values for the pseudoexperiment.  
 From this, expected 95\% credibility level (CL) upper limits 
on cross section times branching ratio are extracted.

\begin{figure}[h!]
  \begin{center}
    \includegraphics[width=0.35\textwidth, clip=true, viewport=0.7in 0.4in 7.8in 5.9in] 
       {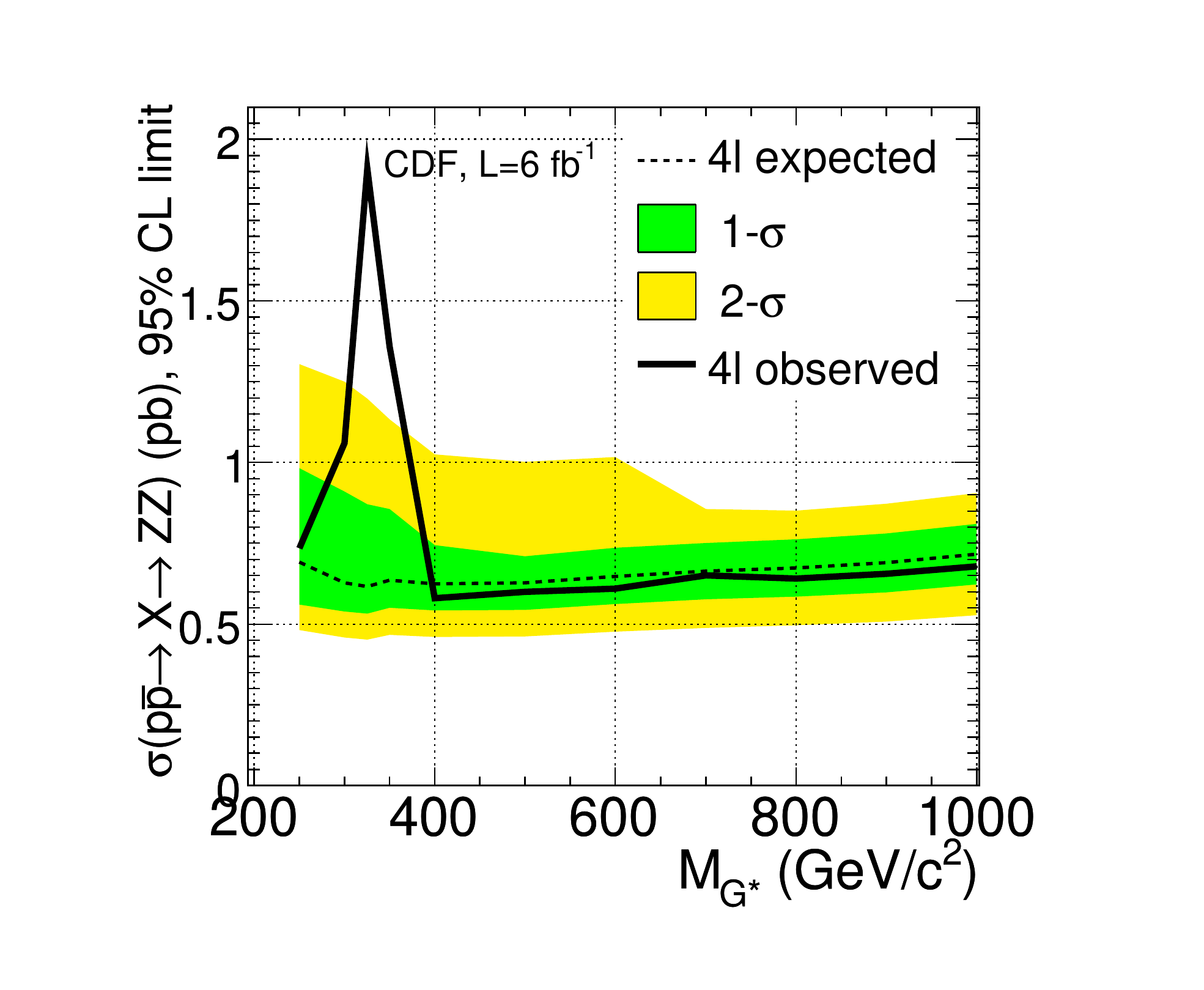}
       \caption[]{Expected and observed 95\% CL limits on $\sigma(\ppbar \to G^* \to ZZ)$ from the four-lepton channel; the four events with \mzz=327\gevcsq\ result in a deviation from the expected limit.}
    \label{fig:bands_zz4l_expected_and_observed}
  \end{center}
\end{figure}

Fig.~\ref{fig:bands_zz4l_expected_and_observed} shows 
expected and observed limits in the four-lepton channel 
for $G^*$ masses between 250 and 1000\gevcsq.
At \mgstar\ the expected sensitivity 
is around 0.7\pb, and the four events with masses clustered 
around that value
result in an observed limit of 1.9\pb.

\begin{figure}[h!]
  \begin{center}	
   \includegraphics[width=0.22\textwidth, clip=true, viewport=0.68in 0.4in 6.85in 6.0in] 
       {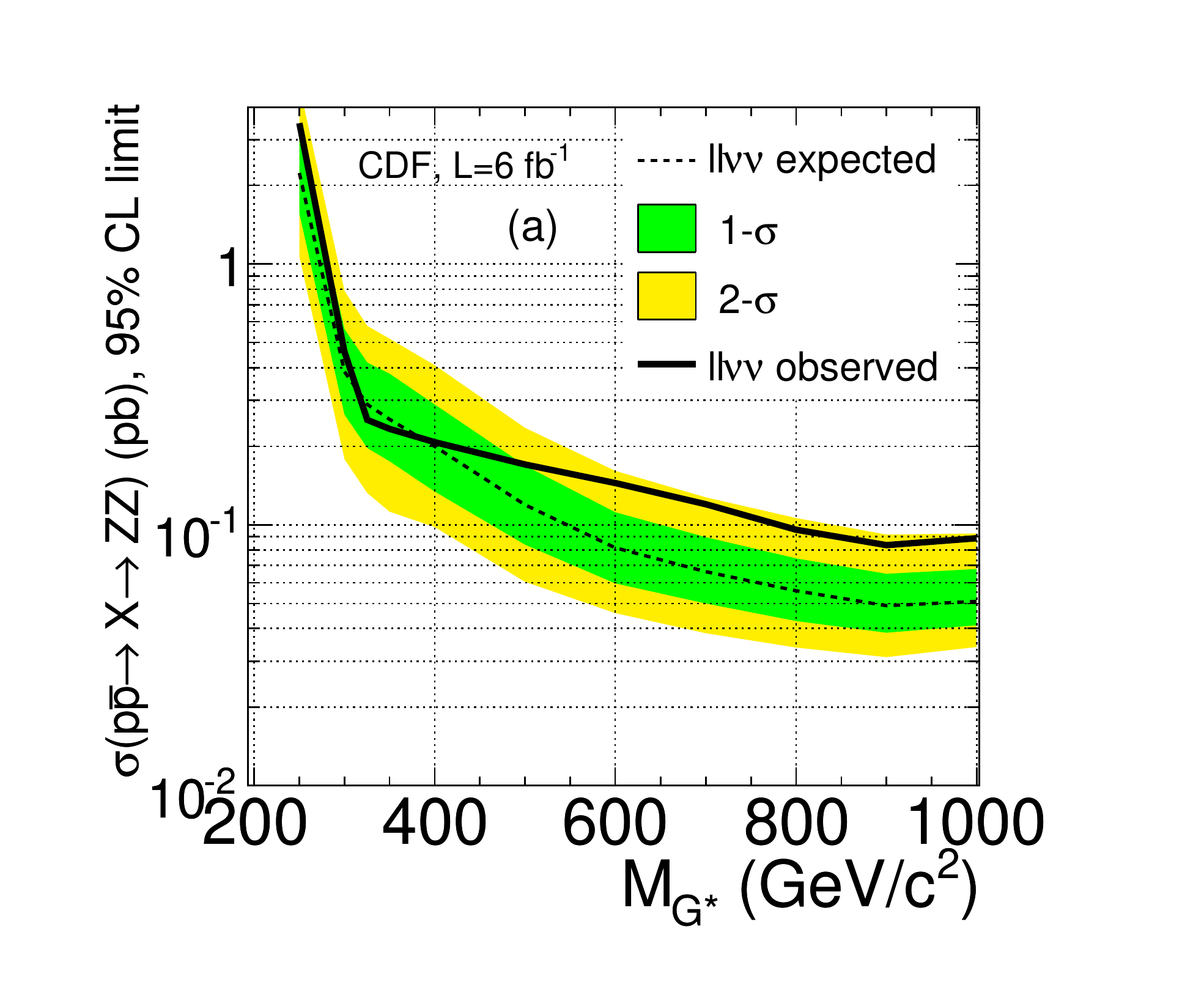}
    \includegraphics[width=0.22\textwidth, clip=true, viewport=0.68in 0.4in 6.85in 6.0in]     
       {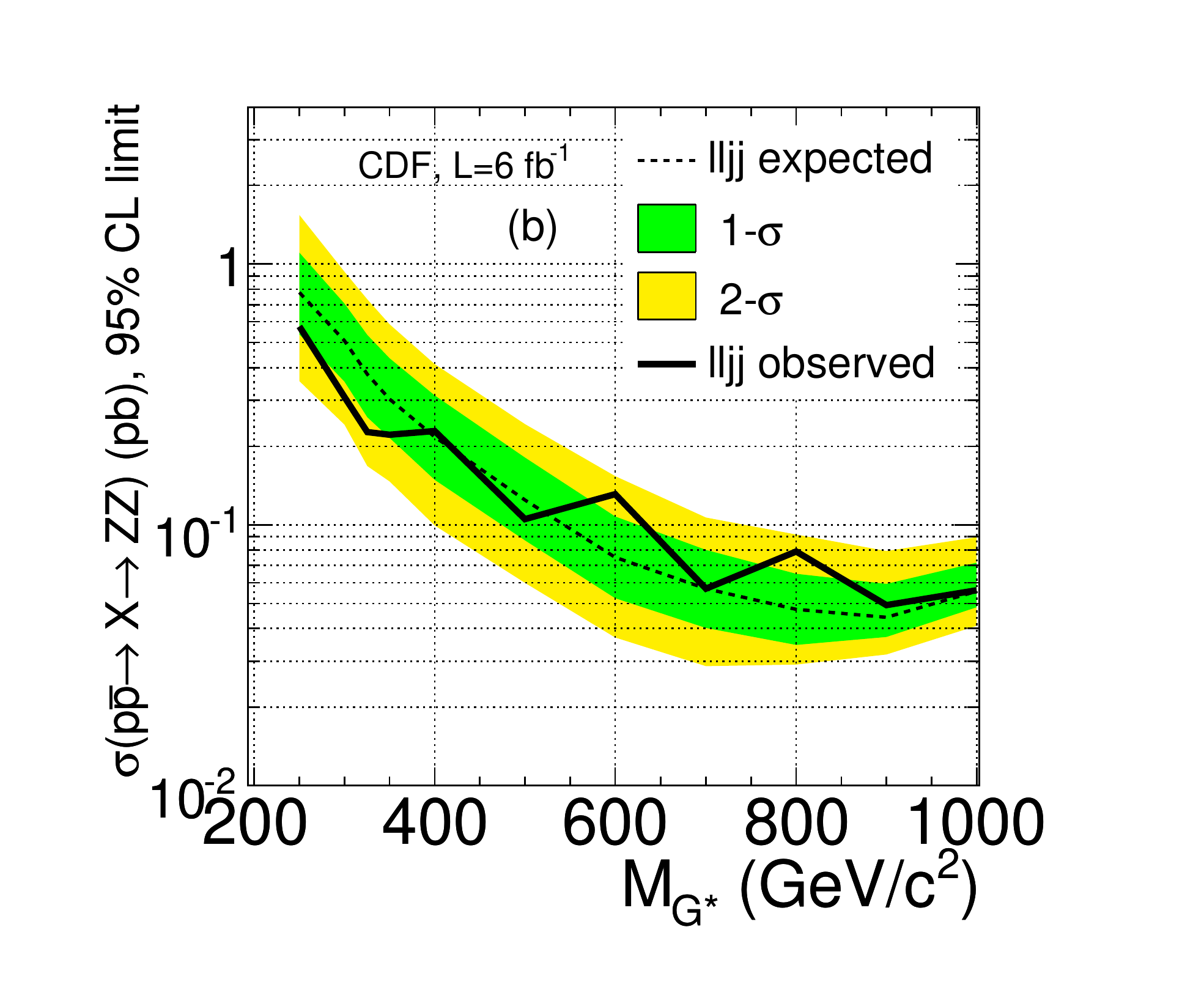}
       \caption[]{Expected and observed 95\% CL limits on $\sigma(\ppbar \to G^* \to ZZ)$ from (a) the \zzllnn\ channel, and (b) the \zzlljj\ channel. }
    \label{fig:otherlimits}
  \end{center}
\end{figure}

Although the backgrounds in the 
\llmet\ channel are higher than in the four-lepton channel, 
this channel provides better sensitivity.
Fig.~\ref{fig:otherlimits}(a) shows the expected 
and observed cross section limits for \llmet, and 
there are no large differences from standard model expectations.
For \mgstar\ the expected 95\% CL upper cross section 
limit on the $s$-channel signal model 
is 0.29\pb, and the observed limit is 0.25\pb.
For the boosted $G^*$ signal model 
the 95\% CL expected and observed limits are both 0.30\pb.
This is a change of less than 10\% from the $s$-channel model,
demonstrating that the analysis 
sensitivity is not strongly dependent on the 
detail of the production model.

Fig.~\ref{fig:otherlimits}(b) shows the expected 
and observed cross section limits for the \lljj\ channel.
Here the expected 95\% CL
upper cross section limit is 0.38\pb\ for \mgstar, 
and the observed limit is 0.23\pb.
With the selection tuned
 for a boosted signal model, $\pt(\lljj)>40$\gevc,
 the sensitivity is improved slightly compared 
to the $s$-channel signal model.
The expected limit is 0.27\pb\ and the 
observed limit is 0.26\pb, showing that also 
in this channel the analysis sensitivity is not strongly 
dependent on the detail of the signal model.

\begin{figure}[h!]
  \begin{center}
    \includegraphics[width=0.32\textwidth, clip=true, viewport=0.7in 0.4in 7.8in 5.9in] 
    {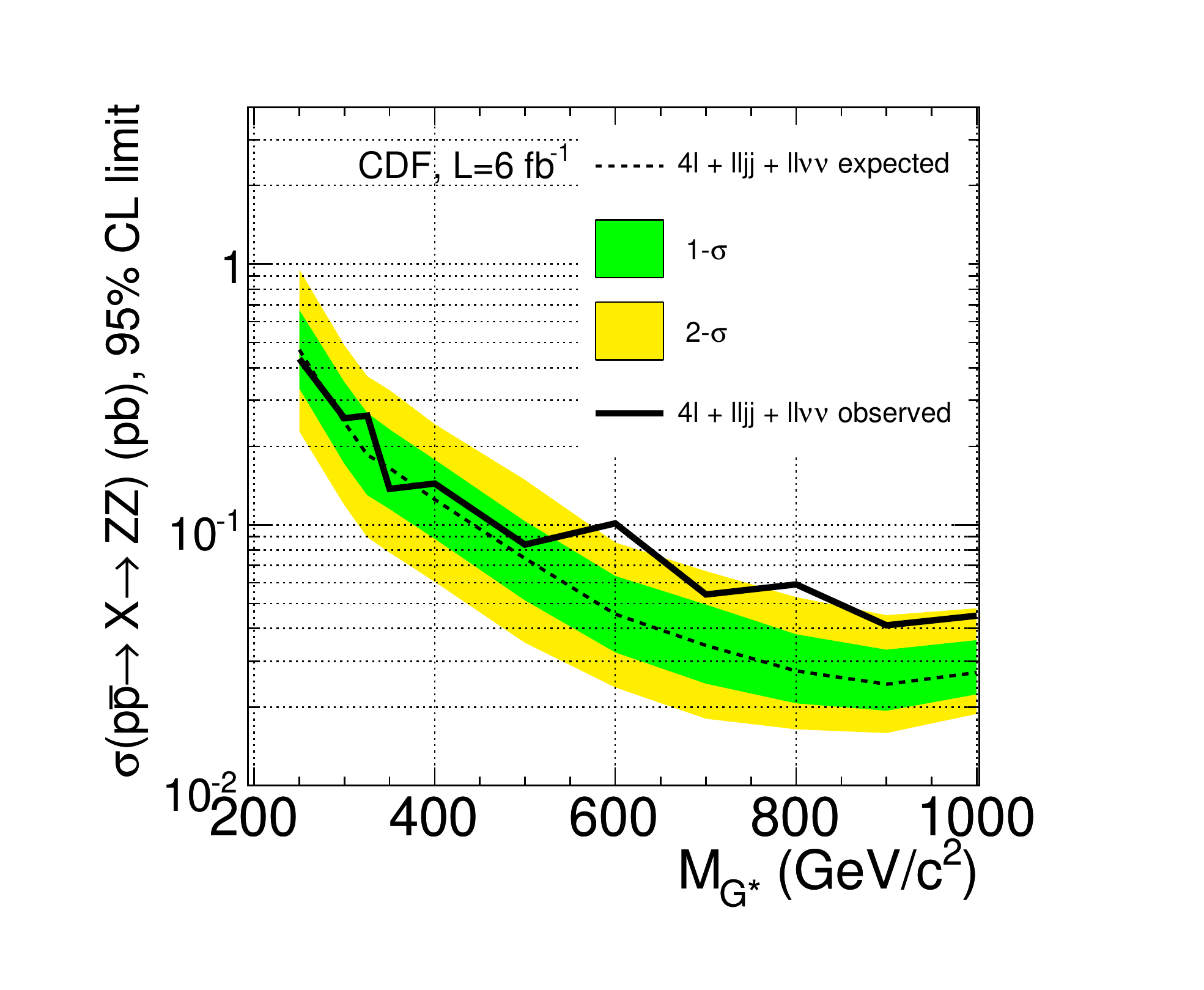}
       \caption[]{Expected and observed 95\% CL limits on $\sigma(\ppbar \to G^* \to ZZ)$ from all channels combined. }
    \label{fig:bands_comb_expected_and_observed}
  \end{center}
\end{figure}

Combining all three channels results in the most sensitivity. 
Expected and observed 
limits are consistent with each other, as shown in 
Fig.~\ref{fig:bands_comb_expected_and_observed}.  For \mgstar\ 
the sensitivity is dominated by the \llmet\ channel.
For an $s$-channel resonance, the 95\% CL upper cross section 
limit is expected to be 0.19\pb\ and is observed to be 0.26\pb.
For a boosted resonance of \mgstar\ the expected limit is 
0.17\pb\ and the observed limit is 0.28\pb.

\section{Conclusions}

We have searched for heavy resonances decaying into \z\ 
boson pairs using the final states consisting of four leptons, 
two leptons and \met, and two leptons plus jets. 
In the channel with the smallest background, the four-lepton channel,
 we have observed eight candidate events.
Four events with high values of \zz\ mass are close in mass, 
and two of those have unusually high $p_T(\zz)$.

However, more sensitive searches in the \llmet\ and \lljj\ 
final states show no indication 
of a new heavy particle 
decaying to two \z\ bosons, suggesting that the events 
observed around $325$\gevcsq\ in the four-lepton channel 
result from standard model processes.
Combining all three channels 
we set upper limits on the cross section 
times branching ratio $\sigma(\ppbar\to G^* \to\zz)$
that vary between 0.26\,pb and 0.045\,pb
  in the mass range $300<M_{G^*}<1000\gevcsq$, 
and the limits do not depend strongly on the production model.

We thank the Fermilab staff and the technical staffs of the
participating institutions for their vital contributions. 
This work was supported by the U.S. Department of Energy and National Science Foundation; the Italian Istituto Nazionale di Fisica Nucleare; the Ministry of Education, Culture, Sports, Science and Technology of Japan; the Natural Sciences and Engineering Research Council of Canada; the National Science Council of the Republic of China; the Swiss National Science Foundation; the A.P. Sloan Foundation; the Bundesministerium f\"ur Bildung und Forschung, Germany; the Korean World Class University Program, the National Research Foundation of Korea; the Science and Technology Facilities Council and the Royal Society, UK; the Russian Foundation for Basic Research; the Ministerio de Ciencia e Innovaci\'{o}n, and Programa Consolider-Ingenio 2010, Spain; the Slovak R\&D Agency; the Academy of Finland; and the Australian Research Council (ARC).



\begin{thebibliography}{99}


\bibitem{higgsxs}
D. de Florian and M. Grazzini, Phys. Lett. B {\bf 674}, 291 (2009).\\ 
C. Anastasiou, R. Boughezal, and F. Petriello, J. High Energy Phys. {\bf 0904}, 003 (2009).\\ 
A. Djouadi, J. Kalinowski, and M. Spira, Comp. Phys. Commun. {\bf 108 C}, 56 (1998). 


\bibitem{ZZ_EXTRA_DIMENSIONS}
M. Kober, B. Koch, and M. Bleicher, Phys. Rev. D {\bf 76}, 125001 (2007).

\bibitem{rsgraviton}
L. Randall and R. Sundrum, Phys. Rev. D {\bf 83}, 4690 (1999).

\bibitem{k_over_mPl}
The coupling must be large enough to be consistent with the 
apparent weakness of gravity but small enough to prevent the 
theory from becoming non-perturbative.  A natural choice is
$k/M_{Pl}=0.1$, where $k$ is a curvature parameter and $M_{Pl}$ 
is the Planck scale.

\bibitem{previousresults}
 T. Aaltonen {\it et al.} (CDF Collaboration), Phys.\ Rev.\ Lett.\  {\bf 107}, 051801 (2011). \\ 
 V. Abazov {\it et al.} (D0 Collaboration), Phys. Lett. B {\bf 695}, 88 (2011).\\ 
 G. Aad {\it et al.} (ATLAS Collaboration), Phys. Lett. B {\bf 700}, 163 (2011). \\ 
 S. Chatrchyan {\it et al.} (CMS Collaboration), J. High Energy Phys. {\bf 1105}, 093 (2011).  

\bibitem{bulk}
K. Agashe, H. Davoudiasl, G. Perez, and A. Soni, Phys. Rev. D {\bf 76}, 036006 (2007).  

\bibitem{lisafitzpatrick}
 L. Fitzpatrick, J. Kaplan, L. Randall, and L. Wang, J. High Energy Phys. {\bf 0709}, 013 (2007). 

\bibitem{antonio}
 T. Aaltonen {\it et al.} (CDF Collaboration), Phys. Rev. D {\bf 83}, 112008 (2011).

\bibitem{cdftdr} 
  R. Blair {\it et al.} (CDF Collaboration), FERMILAB-Pub-96/390-E.

\bibitem{silicon}
A. Sill {\it et al.}, Nucl. Instrum. Methods Phys. Res. A {\bf 447}, 1 (2000).

\bibitem{cot}  
  T. Affolder {\it et al.},  Nucl. Instrum. Methods Phys. Res. A {\bf 526}, 249 (2004).

\bibitem{cdf_cem_ces} 
  L. Balka {\it et al.}, Nucl. Instrum. Methods Phys. Res. A {\bf 267}, 272 (1988).

\bibitem{cdf_cha} 
  S. Bertolucci {\it et al.}, Nucl. Instrum. Methods Phys. Res. A {\bf 267}, 301 (1988).

\bibitem{cdf_muon_system} 
  G. Ascoli {\it et al.}, Nucl. Instrum. Methods Phys. Res. A {\bf 268}, 33 (1988).


\bibitem{clc} 
  D. Acosta {\it et al.}, Nucl. Instrum. Methods Phys. Res. A {\bf 494}, 57 (2002).

\bibitem{drell-yan} 
  A. Abulencia {\it et al.} (CDF Collaboration), J. Phys. G Nucl. Part. Phys. \textbf{34}, 2457 (2007).


\bibitem{run1wz} 
  F. Abe {\it et al.} (CDF Collaboration), Phys. Rev. D \textbf{44}, 29 (1991).


\bibitem{jetclustering}
  F. Abe {\it et al.} (CDF Collaboration), Phys. Rev. D {\bf 45}, 1448 (1992). 

\bibitem{jetcorr}
  A. Bhatti {\it et al.}, Nucl. Instrum. Methods A {\bf 566}, 375 (2006).

\bibitem{pythia} 
  T. Sj\"{o}strand {\it et al.}, Comput. Phys. Commun. {\bf 135}, 238 (2001).

\bibitem{cdfsim}
 E. Gerchtein and M. Paulini,  CHEP-2003-TUMT005.

\bibitem{mcnlo}
  S. Frixione and B. R. Webber, J. High Energy Phys. {\bf 0206}, 029 (2002).

\bibitem{herwig}
 G. Corcella {\it et al.}, J. High Energy Phys. {\bf 01}, 010 (2001).

\bibitem{pdg}
K. Nakamura et al. (Particle Data Group), J. Phys. G \textbf{37}, 7A (2010).

\bibitem{NLO_CROSS_SECTIONS}
J. M. Campbell and R. K. Ellis, Phys. Rev. D {\bf 60}, 113006 (1999). \\
N. Kidonakis and R. Vogt, Phys. Rev. D \textbf{78}, 074005 (2008). 

\bibitem{feldmancousins}
  G. J. Feldman and R. D. Cousins, Phys. Rev. D {\bf 57}, 3873 (1998).

\bibitem{matteo}
CDF Collaboration, to be submitted to Phys.\ Rev.\ Lett.; reference will be updated.

\bibitem{alpgen}
  M. L. Mangano {\it et al.}, J. High Energy Phys. {\bf 0307}, 001 (2003).

\bibitem{joel} 
  J. Heinrich and L. Lyons, Ann. Rev. of Nucl. and Particle Sci. \textbf{57}, 145 (2007).

\bibitem{junk} 
  T. Junk, Nucl. Instrum. Methods Phys. Res. A {\bf 434}, 435 (1999).


\end{thebibliography}
\end{document}